\documentclass[12pt]{article}
\usepackage{latexsym,amsmath,amscd,amssymb,framed,graphics,mathrsfs,diagrams}
\usepackage[colorlinks]{hyperref}
\usepackage{url}
\usepackage[all]{xy}

\def\contract{\makebox[1.2em][c]{\mbox{\rule{.6em}
{.01truein}\rule{.01truein}{.6em}}}}

\newcommand{\rem}[1]{}
\newcommand{\bfi}{\bfseries\itshape}
\newcommand{\bz}{{\bf z}}

\newcommand{\pp}[2]{\frac{\partial #1}{\partial #2}}

\newcommand{\dede}[2]{\frac{\delta #1}{\delta #2}}

\newcommand{\sym}{{\rm sym}}
\newcommand{\bzeta}{\boldsymbol{\zeta}}

\newcommand{\de}{{\mathrm{d}}}

\newcommand{\bw}{{\mathbf{w}}}
\newcommand{\R}{{\mathbb{R}}}

\newcommand{\todo}[1]{\vspace{5 mm}\par \noindent
\framebox{\begin{minipage}[c]{0.95 \textwidth}
\tt #1 \end{minipage}}\vspace{5 mm}\par}

\newtheorem{theorem}{Theorem}[section]

\newtheorem{lemma}[theorem]{Lemma}
\newtheorem{remark}[theorem]{Remark}
\newtheorem{proposition}[theorem]{Proposition}
\newtheorem{corollary}[theorem]{Corollary}

\begin{document}

\title{Vlasov moment flows and geodesics on the Jacobi group}
\author{Fran\c{c}ois Gay-Balmaz$^1$ and Cesare Tronci$^2$\\
\footnotesize\it
$^1$Laboratoire de M\'et\'eorologie Dynamique, \'Ecole Normale Sup�erieure/CNRS, Paris, France.\\
\footnotesize\it
$^2$Section de Math\'ematiques, \'Ecole Polytechnique
F\'ed\'erale de Lausanne, Switzerland}

\date{}

\maketitle

\abstract{By using  the moment algebra of the Vlasov kinetic
equation, we characterize the integrable Bloch-Iserles system on
symmetric matrices \cite{BlBrIsMaRa09} as a geodesic flow on the
Jacobi group
${\rm Jac}(\Bbb{R}^{2n})={\rm Sp}(\Bbb{R}^{2n})\,\circledS\,{\rm H}(\Bbb{R}^{2n})$.
We analyze the corresponding Lie-Poisson structure by presenting a
momentum map, which both untangles the bracket structure and
produces particle-type solutions that
are inherited from the Vlasov-like interpretation.
Moreover, we show how the Vlasov moments associated to Bloch-Iserles dynamics correspond to particular subgroup inclusions
into a group central extension (first discovered by \cite{IsLoMi2006}), which in turn underlies Vlasov kinetic theory.
In the most general case of Bloch-Iserles dynamics, a generalization of the Jacobi group also emerges naturally.}

\tableofcontents

\section{Introduction}

Recent developments on kinetic equations of Vlasov type have shown how different integrable systems (both finite and infinite dimensional) may emerge from appropriate moment closures of the corresponding Vlasov dynamics \cite{HoTr2009}. For example, this is the case for two integrable PDE's in the Camassa-Holm hierarchy \cite{CaHo1993,HoIv2011a,HoIv2011b}. The relation between these integrable systems and the underlying Vlasov kinetic description resides in the fundamental fact that moments are momentum maps which then preserve the geometric structure of the basic kinetic system \cite{HoTr2009}. For example, the Klimontovich particle solution is also a momentum map that is preserved in the case of the Camassa-Holm (CH) equation and this produces the particle description of the well-known peakon solution of the CH equation \cite{CaHo1993}. Analogous arguments apply to a finite-dimensional integrable flow on the space of symmetric matrices, recently proposed by Bloch and Iserles \cite{BlIs06}. 

This paper uses the Vlasov approach to provide a full characterization of the Bloch-Iserles system in terms of its geometric Lie-symmetry properties.
After reviewing the main properties of Vlasov-like equations, this Introduction continues by describing shortly how CH-type systems emerge by taking moments of the Vlasov distribution. Later, the attention is devoted to a special type of moment algebra initially appeared in the context of accelerator beam physics. This moment Lie algebra provides the fundamental understanding of the Bloch-Iserles system, whose main properties are also reviewed in the last part of this Introduction.

\subsection{Kinetic equations of Vlasov type}

Vlasov-type equations govern the evolution of collisionless multi-particle systems that are far from their thermodynamic equilibrium. This is a typical situation in plasma physics, which is the natural context of the Vlasov equation for electrostatic (Poisson-Vlasov equation) or electromagnetic (Maxwell-Vlasov system) interactions. The dynamical variable is a probability distribution on phase space $f(\mathbf{q},\mathbf{p},t)$, which retains the entire statistical information on the system. In the simple case of electrostatic interactions, the Poisson-Vlasov equation maybe written as
\begin{equation}\label{Vlasov}
\pp{f}{t} +\left\{f,\dede{H}{f}\right\}_{can}=0
\end{equation}
relative to the Hamiltonian
\begin{equation}\label{PoissonVlasov}
H(f)=\frac12\int\!f(\mathbf{q},\mathbf{p},t)\left(\left\|\mathbf{p}\right\|^2+U_f(\mathbf{q},t)\right){\rm
d}\mathbf{q}\,{\rm d}\mathbf{p},
\qquad\text{ with } \qquad
\Delta U_f=\int\! f\,{\rm d}\mathbf{p}
\,.
\end{equation}
Here $\left\{\cdot,\cdot\right\}_{can}$ is the canonical Poisson
bracket on $\Bbb{R}^{2k}$ and $U_f$ is the nonlinear
collective potential.

 As for its
Hamiltonian structure, equation \eqref{Vlasov} is well known to be a
Lie-Poisson system on the Poisson algebra $\mathcal{F}(\R^{2k})$ of
phase-space scalar functions. More precisely, equation \eqref{Vlasov} is
Hamiltonian (cf. \cite{Gi1981,Mo1982}) relative to the \textit{Lie-Poisson bracket} (see Appendix \eqref{LP+EP} and references therein)
\begin{equation}\label{VlasovPB}
\left\{F,G\right\}_+(f)=\left\langle f,\left\{\frac{\delta F}{\delta
f} ,\frac{\delta G}{\delta f}\right\}_{can}\right\rangle,
\end{equation}
where the pairing
$\langle\cdot,\cdot\rangle:\mathcal{F}^*(\R^{2k})\times\mathcal{F}(\R^{2k})\to\R$
is  the usual $L^2-$pairing  and the dual space
$\mathcal{F}^*(\R^{2k})={\rm Den}(\R^{2k})$ is given by
distributions on phase-space.

The Lie group underlying Vlasov-type equations is often considered to be the group of symplectic transformations on phase space \cite{Ma82}. However, as pointed out in \cite{MaWe1981,MaWeRaScSp1983,ScWe1994}, the rigorous version of this Lie group is given by the group of strict contact transformations
\[
\operatorname{Cont}(\R^{2k+1}, \theta )=\left\{\varphi\in \operatorname{Diff}(\R^{2k+1})\mid
\varphi^*\theta=\theta \right\},
\]
where $\theta ={\bf p}\cdot\mathbf{d} {\bf q}-\mathbf{d}  s$ is the
contact one-form on $\R^{2k+1}$ and $\operatorname{Diff}(\R^{2k+1})$ is the group of diffeomorphisms of $\R^{2k+1}$. At the infinitesimal level, the
identification between the Lie algebra
$\mathfrak{cont}(\R^{2k+1}, \theta )=\left\{X\in \mathfrak{X} ( \R^{2k+1}) \mid \pounds_X \theta =0\right\}$ of strict contact vector fields (here $\pounds$ denotes Lie
derivation and $ \mathfrak{X}( \R^{2k+1})$ is the space of vector fields on $ \R^{2k+1}$) and the Poisson algebra $\mathcal{F}(\Bbb{R}^{2k})$ of phase-pace functions goes back to Van Hove's thesis
\cite{VanHove}, which contains a proof in coordinates.

\begin{remark}[The group of quantomorphisms]\normalfont
The above picture generalizes to the case of a symplectic manifold $(M, \Omega )$ that is prequantizable. In this case there exists a circle bundle ${P\rightarrow M}$ together with a connection $\theta$ of curvature $\Omega $. Strict contact transformations on $P$ (such that $\varphi^*\theta=\theta$) are necessarily automorphisms of the principal bundle. This explains the common notation $ \operatorname{Cont}(P, \theta )= \operatorname{Aut}(P, \theta )$ emphasizing the $\theta$-preserving nature of these automorphisms, which are also well known as \emph{quantomorphisms}. This is the usual setting in
{\it quantization theory}. For  Vlasov-type equations, the most interesting case is when the symplectic form is exact, e.g. $M=T^*Q$. This picture is developed in the last part of the present paper.
\end{remark}

\begin{remark}[The Klimontovich solution]\normalfont
The Vlasov equation exhibits relevant geometric structures that were recently reviewed in \cite{HoTr2009}. In particular, the Vlasov equation allows for a \emph{dual pair} of momentum maps that is identical to the dual pair structure existing for Euler's fluid equation. The left leg of this dual pair is given by the well known Klimontovich solution
\begin{equation}\label{Klimontovich}
f({\bf q,p},t)=\int_S \!w(s)\,\delta({\bf q-Q}(s,t))\,\delta({\bf p-P}(s,t))\,{\rm d}^ms\,,
\end{equation}
where $s$ is a point on a submanifold $S\subset\Bbb{R}^{2k}$ of dimension $m$ (with $m\leq 2k$), endowed with the volume form $w(s)\,{\rm d}^m s$. The above expression is a momentum map which returns the well known Klimontovich  solution (single-particle) \cite{Kl1967} when $\operatorname{dim}S=0$ (see also \cite{Weinstein83}).
\end{remark}

\subsection{Integrable Vlasov moment closures}\label{Sec:IntMomClo}

While the Poisson-Vlasov system mentioned in the previous section is of paramount importance in plasma physics, its underlying geometry lacks a geometric property which is common to many other Lie-Poisson systems, such as Euler's equations for fluids and rigid bodies. Indeed, the flow of the Poisson-Vlasov system is not geodesic since the Hamiltonian \eqref{PoissonVlasov} contains a linear term in $f$. Then one is led to ask about the possibility of a purely quadratic Vlasov Hamiltonian. This question has been recently pursued in a series of papers \cite{GiHoTr2005,GiHoTr2007,HoTr2009}, which showed how several integrable systems emerge naturally by taking moments of a Vlasov distribution undergoing a geodesic flow on ${\rm Cont}( \mathbb{R}  ^{2k+1}, \theta )$. More particularly, a \emph{geodesic Vlasov equation} reads as
\begin{equation}\label{geodesic-Vlasov}
\frac{\partial f}{\partial t}+\Big\{f,\,\mathcal{G}* f\Big\}=0,
\end{equation}
where $f=f(z,t)$ is the Vlasov distribution, while $\mathcal{G} * f$ denotes convolution with a kernel $\mathcal{G}(z,z')$.
In this context, the momentum map properties of the moment operation have the feature of taking the geodesic flow  on ${\rm Cont}( \mathbb{R}  ^{2k+1}, \theta )$ to another geodesic flow whose properties depend on how moments are taken.

The most celebrated moment method involves tensor fields of the type
\[
A_m(q)=\int p^m \,f(q,p)\,{\rm d}p\,.
\]
In the context of integrable systems, the above quantities first emerged in the study of the Benney integrable hierarchy \cite{Be1973,Gi1981}. These moments are called kinetic moments and their momentum map properties were investigated in \cite{GiHoTr2008} together with their Hamiltonian structure, which is Lie Poisson on the Lie algebra given by the Schouten bracket on symmetric tensor fields \cite{Ni1955}. Kinetic moments were shown to provide a general understanding of integrable PDE's of Camassa-Holm type \cite{GiHoTr2007}. This is related to the fact that moments project the group ${\rm Cont}( T^*Q, \theta )$ underlying Vlasov-type systems to its subgroups, e.g. point transformations ${\rm Diff}(Q)$ and their extended semidirect-product version  ${\rm Diff}(Q)\,\circledS\,\mathcal{F}(Q)$ (here $Q$ is
the particle configuration manifold).  For example, the (one
component) Camassa-Holm equation is a geodesic equation on the
diffeomorphism group ${\rm Diff}(\Bbb{R})$. These
transformations are regarded as canonical transformations on the phase
space $\Bbb{R}^2=T^*\Bbb{R}$ that are generated by
functions $h(z)=h(q,p)$ that are (homogeneous) linear in the canonical one-form
${p} \, \mathrm{d} {q} $. This class of linear functions in the canonical momentum is naturally associated to the first-order moment, whose dynamics can recover the Camassa-Holm equation as described in \cite{GiHoTr2005}. This case correspond to  $\mathcal{G}=p\,p' e^{-|q-q'|/\alpha}$. An analogous argument holds for the
two-component Camassa-Holm (CH2) equation \cite{GiHoTr2007}, which is a geodesic flow on
the semidirect-product group ${\rm
Diff}(\Bbb{R})\,\circledS\,\mathcal{F}(\Bbb{R})$. This Lie group is generated by phase space functions that are
(inhomogeneous) linear in the canonical one-form. These functions are naturally associated to both the zero-th and first order moments, whose dynamics returns the two component Camassa-Holm equation \cite{ChLiZh2006} for an appropriate choice of Hamiltonian. Recovering the CH2 system from the geodesic Vlasov equation requires setting $\mathcal{G}=e^{-|q-q'|}+p\,p' e^{-|q-q'|/\alpha}$.

Besides the integrable CH and CH2 equations, the Vlasov approach has been also shown to apply to another finite-dimensional integrable system, which was introduced by Bloch and Iserles \cite{BlIs06} and will be described in the following sections. This system was shown \cite{HoTr2009} to arise from a quadratic Vlasov Hamiltonian upon taking another type of moments, which are different from the kinetic moments mentioned above. These alternative moments will be called \emph{statistical moments} and are presented in the next section. The geodesic flow associated to the Bloch-Iserles system is completely understood only in one special case, in which this system describes a geodesic on the Lie group $\operatorname{Sp}(\Bbb{R}^{2k})$ of symplectic matrices \cite{BlBrIsMaRa09}. However, more general cases are possible for the Bloch-Iserles dynamics and this paper aims to provide a complete characterization of their associated geodesic flows.

\begin{remark}[Momentum map solutions of moment equations]\label{SolMomaps} \normalfont 
Another geometric property common to many moment equations is that they possess special solutions that are also momentum maps. These solutions arise from the remarkable fact that the Klimontovich momentum map \eqref{Klimontovich} survives the process of taking moments \cite{HoTr2009}. Then, for example, the Klimontovich solution yields the particle description of the peakon solution for both the CH equation and a modified version of the CH2 system \cite{HoOnTr2009}. As we shall see, an analogous argument also holds for the Bloch-Iserles system.
\end{remark}

\subsection{Statistical moments and their Lie-Poisson structure}
As mentioned in the previous section, the Vlasov moment interpretation applies to  the Bloch-Iserles system \cite{HoTr2009,BlBrIsMaRa09}, provided moments are taken appropriately. In particular, this approach is based on the definition of \emph{statistical moments}, whose corresponding Lie algebra was studied
in detail by \cite{ScWe1994}. This section reviews the basic properties of these moments in preparation for their application to the Bloch-Iserles system, which in turn is briefly recalled in the next section.

Upon denoting $\bz=({\bf q,p})$, the $n-$th
{statistical moment} is defined by the following symmetric
contravariant tensor on $\R^{2k}$:
\begin{equation}\label{statmoms}
X_m(f):=\frac1{m!}\int \bz^{\otimes m}f(\bz)\ {\rm d}\bz\
\in\,\bigvee_{i=0}^m\,\Bbb{R}^{2k},\quad m=0,1,2,...\,,
\end{equation}
where `$\vee$' denotes the symmetrized tensor product $\otimes^{\rm
sym}$, and $X_0$ is the (constant) total number of particles:
usually one fixes the normalization $X_0=1$, although we shall
consider a generic value of $X_0$. 
\begin{remark}[Statistical moments of accelerator beams]\normalfont 
Statistical moments were introduced by \cite{Ch1983}, in the context of beam dynamics for particle accelerators. In accelerator beam optics, the dynamics of the first three moments governs the linear dynamics of particle beams moving through a magnet lattice. When nonlinear effects are also taken into account, higher order moments must be considered and the moment hierarchy must be truncated properly \cite{Ch1995}. Moment (Casimir) invariants were classified in \cite{HoLySc1990}.
\end{remark}
We shall refer to the quantities
\eqref{statmoms} as simply `moments', unless specified differently.  
These moments take the Vlasov Lie-Poisson structure \eqref{VlasovPB} to the following Lie-Poisson bracket:
\begin{align}\label{momentPB}
\{F,G\}(X) 
&=: \sum_{n,m=0}^\infty \left\langle X_{n+m-2},\left[\frac{\partial F}{\partial
X_n},\frac{\partial G}{\partial X_m}\right]\right\rangle,
\end{align}
where the
moment Lie bracket is defined as
\begin{equation}\label{momLiebrkt}
\left[S_n,S_m'\right]:=-\frac{(n+m-2)!}{(n-1)!\,(m-1)!}\left(S_n\,\Bbb{J}^{-1}
S_m'\right)^{\rm sym}
\end{equation}
and the pairing $\langle\cdot,\cdot\rangle$ is given by tensor
contraction.
Here, $A^{\rm sym}$ denotes the symmetric part of the tensor $A$, while $\Bbb{J}$ is the  Poisson matrix that can be chosen canonical for simplicity:
\begin{equation}\label{canonical_symplectic_structure}
\Bbb{J}= \left(
\begin{array}{cc}
 0 & 1
 \\
-1 & 0
\end{array}
\right)=-\Bbb{J}^{-1}.
\end{equation}
 Also,
$A B$ denotes the one-index contraction between covariant and
contravariant tensors, that is $(A B)_{ij...}^{hl...} =
A_{ij...k}B^{khl...}$ (analogously for $B A=(B
A)^{km...}_{jl...} = B^{km...i}A_{ijl...}$). We may rephrase  the
above considerations  in the following statement:
\begin{theorem}{\rm \cite{HoLySc1990}}
Let
\[
S\mathcal{T}(V):=\bigoplus_{n=0}^\infty\left(\,\bigvee_{i=0}^n\,
V^{*\!}\right)
=:\bigoplus_{n=0}^\infty \,S\mathcal{T}_n(V)
\]
be the Fock space of symmetric covariant tensors on the
symplectic space $(V,\Bbb{J})$, endowed with the Lie bracket
\eqref{momLiebrkt}.
The map
\[
\alpha\,:\,\big(S\mathcal{T}(V),[\cdot,\cdot]\big)\rightarrow \big(\mathcal{F}(V),\{\cdot,\cdot\}\big),\;\;S_n \mapsto\sum \frac1{n!}\ S_n\contract\, \mathbf{z} ^{\otimes\,n}
\]
is a Lie algebra homomorphism.
\end{theorem}

\begin{remark}[Arbitrary constant symplectic structure]\label{remark-on-N}\normalfont
Although the remainder of this section uses the canonical symplectic form, so that $\Bbb{J}=-\Bbb{J}^{-1}$,  the following discussions in this paper often refer to an arbitrary symplectic structure $\mathbb{J}$ on $\mathbb{R}^{2k}$. Then, when $\Bbb{J}$ is not
the canonical structure, the corresponding Poisson tensor is given in matrix form by
\[
\mathbb{N}:=-\mathbb{J}^{-1}
\]
as it appears in the Lie bracket \eqref{momLiebrkt}. The notation $\Bbb{N}$ is chosen to be consistent with \cite{BlBrIsMaRa09}.
\end{remark}

\begin{remark}[Conventions on factorials]\normalfont In the expressions above
one has to recall that $[S_0,S_n']=[S_n,S_0']=0\ \forall n\geq0$, as
it can be seen by expanding polynomials as
\[
\sum_{n\geq0} \frac1{n!}\ S_n\contract\,
\mathbf{z} ^{\otimes\,n}=S_0+\sum_{n\geq1} \frac1{n!}\ S_n\contract\,
\mathbf{z} ^{\otimes\,n}\,.
\]
For the Lie algebra coadjoint operator, this means that ${\rm
ad}^*_{S_0}\,X_{n}={\rm ad}^*_{S_n}\,X_{n-2}=0$, as it is shown
by relabeling the indexes in the moment bracket:
\[
\{F,G\}= \sum_{n,m} \left\langle X_{n+m-2}\,,\left[\frac{\delta
F}{\delta X_n},\frac{\delta G}{\delta X_m}\right] \right\rangle =
\sum_{n,k} \left\langle {\rm ad}^*_{\,\delta F/\delta X_n\,} X_k, \,
\frac{\delta G}{\delta X_{k-n+2}} \right\rangle.
\]
All these cancelations can be incorporated by simply defining
$(-n)!:=0$, which is the convention adopted throughout this paper.
\end{remark}

From the discussion above one can easily recognize the graded
structure of the moment algebra, with filtration
\[
\left[S\mathcal{T}_n(V),S\mathcal{T}_m(V)\right]\subseteq
S\mathcal{T}_{n+m-2}(V)
\]
Then, one can see that the space
$S\mathcal{T}_2(V)$ dual to second order moments is a subalgebra,
i.e. the algebra ${\rm Sym}(V)=V^*\vee V^*$ of quadratic forms on
$V$, where we recall the notation $V^*\vee V^*:=\left(V^*\otimes
V^*\right)^{\rm sym}$. The largest subalgebra is given by ${\rm
Sym}(V)\oplus V^*\oplus\R$ and it will play a central role in the
remainder of this paper. In order to better understand the structure of the subalgebra ${\rm
Sym}(V)\oplus V^*\oplus\R$, one can restrict the moment Lie-Poisson bracket \eqref{momentPB} to consider only the first three moments
\[
X_0(f)=\int f(\bz)\ {\rm d}\bz \,,\qquad X_1(f)=\int \bz f(\bz)\ {\rm
d}\bz\,,\qquad X_2(f)=\frac12\int \bz^{\otimes 2} f(\bz)\ {\rm d}\bz\,.
\]
Upon choosing the canonical symplectic form $\Bbb{J}=-\Bbb{J}^{-1}$, this produces the Lie-Poisson bracket
\begin{align}\label{mombrkt_symtens}
\left\{F,G\right\}_+(X_2,X_1,X_0)=&\operatorname{Tr}\left(X_2\left[\frac{\partial F}{\partial X_2},\frac{\partial G}{\partial X_2}\right]_{\mathbb{J}}\right)\nonumber\\
&+X_1^{\,T\!}\left(\frac{\partial F}{\partial X_1}\mathbb{J}\frac{\partial
G}{\partial X_2} -\frac{\partial G}{\partial
X_1}\mathbb{J}\frac{\partial
F}{\partial X_2}\right)\\
&+ X_{0\,} \frac{\partial F}{\partial X_1}\mathbb{J}\left(\frac{\partial
G}{\partial X_1}\right)^T.
\nonumber
\end{align}
For vanishing \makebox{$(X_0,X_1)$}, the above Poisson bracket
 reduces to Lie-Poisson dynamics on the Lie
algebra ${\rm Sym}(\Bbb{R}^{2k})$ of symmetric matrices on
phase-space, with Lie bracket given by
\[
\left[S_2,S_2'\right]_{\mathbb{J}}=S_{2\,}\Bbb{J}S_2'-S'_{2\,}\Bbb{J}S_2
\,.
\]
This Lie algebra is generated
by homogeneous quadratic polynomials on phase space, which are well
known to arise as the Lie algebra of the  symplectic group
${\rm Sp}(\Bbb{R}^{2k})$ \cite{GuSt1984}. On the other hand, the
space of inhomogeneous quadratic polynomials (i.e. non-vanishing $(
X_0,X_1)$) is known in quantization theory as the Lie
algebra of the Jacobi group \cite{GuSt1984}
\[
{\rm Jac}(\Bbb{R}^{2k})={\rm Sp}(\Bbb{R}^{2k})\ \circledS\ {\rm H}(\Bbb{R}^{2k}),
\]
where H$(\Bbb{R}^{2k})$ is the Heisenberg group on
$\Bbb{R}^{2k}\times\Bbb{R}$. The semidirect-product structure (denoted \makebox{by $\circledS$}) of
the Jacobi group has attracted a certain attention over the last
decades in the quantization community. For applications of the Jacobi group in quantum mechanics see \cite{Berceanu2007,Berceanu2007b,Berceanu2008,Berceanu2009}. Also,  applications of the Jacobi group to signal processing were recently proposed in \cite{Shuman2003}.

Remarkably, the moment Poisson bracket \eqref{mombrkt_symtens} applies to the Bloch-Iserles system without substantial modifications, as shown in \cite{HoTr2009}. The present
paper focuses on this particular integrable system thereby
determining its underlying symmetry group in terms of the
corresponding Vlasov moment formulation. This construction involves
a {\it quadratic} moment Hamiltonian, whose associated geodesic flow will be characterized in the following sections.

\begin{remark}[Phase-space averages and momentum maps]\label{Averages}\normalfont
The moment method is used in kinetic theory to take averages of microscopic quantities. This process possesses a particularly interesting meaning in terms of momentum maps. Indeed, a general statement \cite{MaRa99} ensures that any equivariant momentum map ${\bf J}:\Bbb{R}^{2k}\to\mathfrak{g}^*$ (generated by the symplectic action of a Lie group $G$ with Lie algebra $\mathfrak{g}$) produces a momentum map $\mathcal{J}:\operatorname{Den}(\Bbb{R}^{2k})\to\mathfrak{g}^*$ via the relation
\[
\left\langle\mathcal{J}(f),\xi\right\rangle=\int f({\bf q,p})\left\langle{\bf J}({\bf q,p}),\xi\right\rangle\,{\rm d}^k{\bf q}\,{\rm d}^k{\bf p}\,.
\]
Consequently, we can state that
\begin{center}\it
any matrix Lie-Poisson system can be realized as a Vlasov moment equation of some kind,
\end{center}
as long as the $G$-action on $\Bbb{R}^{2k}$ admits an equivariant momentum map. For instance, any $G$-action on $\Bbb{R}^{k}$ can be lifted to a canonical $G$-action on $\Bbb{R}^{2k}=T^*\Bbb{R}^{k}$, which in turn produces always an equivariant momentum map.
A particular example of this class is given by the relation $\mathcal{J}(f)=\int {\bf q\times p}\ f({\bf q,p})\,{\rm d}^k{\bf q}\,{\rm d}^k{\bf p}$, which
takes the Vlasov Lie-Poisson structure to the rigid body Lie-Poisson bracket on $\mathfrak{so}^*(3)$ (i.e. the dual space of antisymmetric matrices). All the above still holds upon replacing the vector phase space $\Bbb{R}^{2k}$ by a finite-dimensional symplectic manifold $(\mathcal{P},\Omega)$ acted on by $G$, e.g. $\mathcal{P}=T^*Q$ for some configuration manifold $Q$.

It is now clear that statistical moments can be considered as arising from averaging of the momentum map
\[
\mathbf{J}:V\to\mathcal{ST}(V)\,,\qquad\mathbf{J}(\bz)=\left\{\frac1{n!}\,\bz^{\otimes n}\right\}_{\!n\in\mathbb{N}}
\]
while kinetic moments arise in the same context from the momentum map \cite{GiHoTr2005,GiHoTr2007}
\[
\mathbf{J}:T^*Q\to\mathcal{ST}^*(Q)\otimes \mathrm{Den}(Q)\,,\qquad\mathbf{J}(\mathbf{q,p})=\left\{\mathbf{p}^{\otimes n}\right\}_{\!n\in\mathbb{N}}\,\delta(\mathbf{q-x}),
\]
where $\mathcal{ST}(V)$ are the contravariant tensors on the symplectic space $V$, while $\mathcal{ST}^*(Q)$ denotes the space of covariant tensor fields on the configuration space $Q$.
\end{remark}

\subsection{The Bloch-Iserles system}

The Bloch-Iserles system \cite{BlIs06,BlBrIsMaRa09} is the ordinary
differential equation on the space ${\rm Sym}(\Bbb{R}^n)$
 of real $n$-dimensional symmetric matrices given by
\begin{equation}\label{BI}
\dot{X}(t)=\left[X(t)^2,\Bbb{N}\right],
\end{equation}
where $\Bbb{N}$ is a constant $n$-dimensional antisymmetric matrix and $[\cdot,\,\cdot]$ denotes the matrix commutator. This system is known to be Lie-Poisson on the space ${\rm Sym}(\Bbb{R}^n)$ endowed with the
Lie bracket
\begin{equation}\label{def_N_bracket}
\left[S,S'\right]_\Bbb{N}:=S\Bbb{N}S'-S'\Bbb{N}S,\quad\text{for all}\quad  S,S'\in{\rm
Sym}(\Bbb{R}^n),
\end{equation}
with Hamiltonian
\begin{equation}\label{BI_Ham}
h(X)=\frac12{\rm Tr}(X^2).
\end{equation}
To see this, we identify ${\rm Sym}(\Bbb{R}^n)$ with its dual via the positive
definite inner product $\langle S,X\rangle:= {\rm Tr}(SX)$, for
$S,X\in {\rm Sym}(\Bbb{R}^n)$. Then, relative to this pairing, one
obtains the infinitesimal coadjoint representation
\[
\operatorname{ad}^*_SX=XS\mathbb{N}-\mathbb{N}SX.
\]
Since $\delta h/\delta X=X$, the (left) Lie-Poisson equation
\eqref{LP} associated to the Lie algebra
$(\operatorname{Sym}(n),[\,,]_{\mathbb{N}})$ and the Hamiltonian $h$
recovers the Bloch-Iserles system \eqref{BI}. The system can also be viewed as an Euler-Poincar\'e
equation, associated to the Lagrangian
\[
\ell(S)=\frac{1}{2}\operatorname{Tr}(S^2).
\]

The Lie group underlying the geometry of the Bloch-Iserles system
will be determined in Section \ref{symmetry_group_BI} below.
We now
recall from \cite{BlIs06} and \cite{BlBrIsMaRa09} the principal
properties of the Bloch-Iserles system.

\paragraph{Lie algebra structure.}
Suppose that the the antisymmetric matrix $\mathbb{N}$ has corank $d$. We can thus write
\[
\mathbb{N}=\left[
\begin{array}{cc}
\bar{\Bbb{N}} & 0
\\
0 & 0
\end{array}
\right],
\]
where $\bar{\Bbb{N}}$ is a $2k\times 2k$ invertible matrix, with
$n=2k+d$. In this case, upon denoting by $\mathcal{M}_{d\times 2k}$
the space of $d\times 2k$ matrices,  we have the Lie algebra
isomorphism
\begin{equation}\label{psi}
\Psi:
\left(\left(\operatorname{Sym}(\mathbb{R}^{2k})\,\circledS\,\mathcal{M}_{d\times
2k}\right)\times_C\operatorname{Sym}(\mathbb{R}^d),[\cdot,\cdot]_C\right)\rightarrow\left(\operatorname{Sym}(\mathbb{R}^n),[\cdot,\cdot]_\mathbb{N}\right)
\end{equation}
given by
\[
\Psi(S,A,B):=\left[
\begin{array}{cc}
S & A^T
\\
A & 2B
\end{array}
\right],
\]
with
\begin{equation}\label{BILieAlg}
\left[(S,A,B),(S', A',
B')\right]_C=\left(\left[S,S'\right]_{\bar{\mathbb{N}}\,},\,A\bar{\mathbb{N}}
S'-A'\bar{\mathbb{N}}S\,,
\left(A\bar{\mathbb{N}}{A}^{\prime\,T}\right)^{\rm sym\,}\right),
\end{equation}
see Proposition 2.4 in \cite{BlBrIsMaRa09} (note that we use
here slightly different conventions).

We now describe the Lie
algebra structure of the left hand side of the isomorphism. The Lie
bracket on $\operatorname{Sym}(\mathbb{R}^{2k})$ is
$[\cdot,\cdot]_{\bar{\mathbb{N}}}$ defined in \eqref{def_N_bracket}, with $\mathbb{N}$ replaced by $\bar{\mathbb{N}}$. The semidirect product Lie algebra $\operatorname{Sym}(\mathbb{R}^{2k})\,\circledS\, \mathcal{M}_{d\times
2k}$ is associated to the Lie algebra action of $\operatorname{Sym}(\mathbb{R}^{2k})$ on $\mathcal{M}_{d\times 2k}$ given by
\begin{equation}\label{LieAlgAct}
A\mapsto S\cdot A:=-A\bar{\mathbb{N}}S
\,,
\end{equation}
where one verifies the condition $[S,S']_{\bar{\mathbb{N}}}\cdot
A=S\cdot ( S'\cdot A)- S'\cdot (S\cdot A)$. This semidirect product
Lie algebra is then extended according to the Lie algebra two
cocycle
\begin{equation}\label{cocycle1}
C((S,A),(S', A'))=\frac{1}{2}\left(A\bar{\mathbb{N}}A^{\prime\,T}-
A'\bar{\mathbb{N}}A^T\right)=\left(A\bar{\mathbb{N}}
A^{\prime\,T}\right)^{\sym},
\end{equation}
where $M^{\sym}$ denotes the symmetric part of a
matrix $M$. We denote by $\left(\operatorname{Sym}(\mathbb{R}^{2k})\,\circledS\, \mathcal{M}_{d\times
2k}\right)\times_C\operatorname{Sym}(\mathbb{R}^d)$ the associated central extended Lie algebra and one can check that the cocycle \eqref{cocycle1} leads to the Lie bracket
\eqref{BILieAlg}. Recall that a $W$-valued two cocycle $C$ on a Lie algebra
$\mathfrak{g}$ is defined as a
bilinear antisymmetric map $C:\mathfrak{g}\times\mathfrak{g}\to W$
such that
\[
C(\left[\xi,\eta\right],\zeta)+C(\left[\zeta,\xi\right],\eta)+C(\left[\eta,\zeta\right],\xi)=0\,,
\quad\ \forall\ \xi,\eta,\zeta\in\mathfrak{g}.
\]

On the Lagrangian side, the isomorphism $\Psi$ yields the
Euler-Poincar\'e Lagrangian of the Bloch-Iserles system
\begin{equation}\label{BI_Lag_Psi}
l(X)=\frac12\operatorname{Tr}(X^2)=\frac{1}{2}\operatorname{Tr}(S^2)+\operatorname{Tr}(A^TA)+2\operatorname{Tr}(B^2)=:\ell(S,A,B).
\end{equation}
We shall see later how this Lagrangian yields an invariant metric on
an appropriate matrix Lie group.

One can also find the corresponding
Lie-Poisson Hamiltonian by simply considering the dual map to $\Psi$
in order to write the correct Lie-Poisson formulation on the dual
Lie algebras. Upon using the trace pairing to identify ${\rm
Sym}^*(\R^n)\simeq{\rm Sym}(\R^n)$, we obtain the map
\[
\Psi^*:\operatorname{Sym}^*(\R^n)\rightarrow
\operatorname{Sym}(\R^{2k})\times\mathcal{M}_{2k\times
d}\times\operatorname{Sym}(\R^d),
\]
given by
\[
\Psi^*\left[
\begin{array}{cc}
Y & K/2
\\
K^T/2 & M/2
\end{array}
\right]=(Y,K,M).
\]
Indeed, we have the equalities
\begin{align*}
\left\langle (Y,K,M),(S,A,B)\right\rangle&=\operatorname{Tr}(YS)+\operatorname{Tr}(KA)+\operatorname{Tr}(MB)\\
&=\left\langle \left[
\begin{array}{cc}
Y & K/2
\\
K^T/2 & M/2
\end{array}
\right],\left[
\begin{array}{cc}
S & A^T
\\
A & 2B
\end{array}
\right]\right\rangle\\
&=\left\langle \left[
\begin{array}{cc}
Y & K/2
\\
K^T/2 & M/2
\end{array}
\right],\Psi(S,A,B)\right\rangle.
\end{align*}
Thus, the
Bloch-Iserles Hamiltonian \eqref{BI_Ham} on
$\operatorname{Sym}(\R^{2k})\times\mathcal{M}_{2k\times
d}\times\operatorname{Sym}(\R^{d})$ reads
\begin{equation}\label{BI_Ham_Psi}
h(X)=\frac12\operatorname{Tr}(X^2)=\frac{1}{2}\operatorname{Tr}(Y^2)+\frac{1}{4}\operatorname{Tr}(K^TK)+\frac{1}{8}\operatorname{Tr}(M^2)=:{\sf
h}(Y,K,M).
\end{equation}
We now consider two particular cases that are related to the moment
dynamics associated to the Vlasov system.

\paragraph{Case 1: $\Bbb{N}$ invertible.} When $n=2k$ and the matrix $\Bbb{N}$ is non degenerate, the Lie bracket \eqref{BILieAlg} restricts to its first component. In this particular case, the Lie algebra
$\left(\operatorname{Sym}(2k),[\cdot,\cdot]_\mathbb{N}\right)$ can
be identified with the Lie algebra of Hamiltonian matrices
\[
\mathfrak{sp}(\R^{2k},\Bbb{N}^{-1}):=\left\{\mathcal{S}\in\mathcal{M}_{2k\times2k}\mid
\Bbb{N}^{-1}\mathcal{S}+\mathcal{S}^T\Bbb{N}^{-1}=0\right\}.
\]
The latter is the Lie algebra of the symplectic group
\[
{\rm Sp}(\R^{2k},\mathbb{N}^{-1})=\left\{g\in
\mathcal{M}_{2k\times2k}\mid
g^T\mathbb{N}^{-1}g=\mathbb{N}^{-1}\right\},
\]
and the Lie algebra isomorphism is given by
\begin{equation}\label{Lie_algebra_isom}
\Gamma_{\mathfrak{sp}}:\left({\rm
Sym}(\R^{2k}),\left[\cdot,\,\cdot\right]_\mathbb{N}\right)\rightarrow
\left(\mathfrak{sp}(\R^{2k},\Bbb{N}^{-1}),\left[\cdot,\,\cdot\right]\right),\quad S\mapsto
\Gamma(S):=\mathbb{N}S=\mathcal{S}.
\end{equation}
Thus, if $\Bbb{N}$ is invertible, the Bloch-Iserles system is the
reduction of a geodesic flow of a left invariant metric on the
linear symplectic group ${\rm Sp}(\R^{2k},\Bbb{N}^{-1})$ \cite{BlBrIsMaRa09}. The
Bloch-Iserles Lagrangian on $\mathfrak{sp}(\R^{2k},\Bbb{N}^{-1})$
reads
\[
l(\mathcal{S})=\frac{1}{2}\operatorname{Tr}\left(\mathcal{S}^{T\!\!}\left(-\mathbb{N}^{-2}\right)\mathcal{S}\right),
\]
which corresponds to the following unreduced left invariant Lagrangian
$L:T\operatorname{Sp}(\R^{2k},\Bbb{N}^{-1})\to\R$:
\[
L(g,\dot{g})=l(g^{-1}\dot
g)=\frac12\operatorname{Tr}\!\left(\dot{g}^T\mathcal{G}(g)\,\dot{g}\right),\quad\text{
with }\quad \mathcal{G}(g)=\left(\Bbb{N}^{-1}g^{-1}\right)^{T\!}
\left(\Bbb{N}^{-1}g^{-1}\right)=-g^{T\,}\Bbb{N}^{-2\,} g.
\]
\rem{ 
\todo{C: Any comments about this formula, Francois?\\
F: I obtain a different result:
\[
L(g,\dot{g})=l(g^{-1}\dot g)=\frac12\operatorname{Tr}\!\big((g^{-1}\dot{g})^T(-\Bbb{N}^{-2})g^{-1}\dot{g}\big)=\frac12\operatorname{Tr}\!\big(\dot{g}^T\mathcal{G}(g)\,\dot{g}\big),\qquad\text{
with }\qquad \mathcal{G}(g)=-g^{-T\,}\Bbb{N}^{-2\,} g^{-1}
\]

C: These formulas are equivalent, once we use
$g^T\mathbb{N}^{-1}g=\mathbb{N}^{-1}$.}
}      
In order to write the Lie-Poisson equations, we identify the Lie
algebra $\mathfrak{sp}(\Bbb{R}^n,\Bbb{N}^{-1})$ with its dual via
the Killing form $\langle
\mathcal{X},\mathcal{S}\rangle=\operatorname{Tr}(\mathcal{X}\mathcal{S})$.
One then obtains the coadjoint representation
$\operatorname{ad}^*_\mathcal{S}\mathcal{X}=[\mathcal{X},\mathcal{S}]$
and the \textit{left} Lie-Poisson equations
\[
\dot{\mathcal{X}}=\left[\mathcal{X},\frac{\delta h}{\delta\mathcal{X}}\right].
\]
With respect to the same pairing, the dual map to the Lie algebra isomorphism $\Gamma$ reads
\begin{equation}\label{alpha_star}
\Gamma_{\mathfrak{sp}}^*:\mathfrak{sp}(\R^{2k},\Bbb{N}^{-1})^*\rightarrow {\rm
Sym}(\R^{2k})^*,\quad
\mathcal{X}\mapsto\Gamma_{\mathfrak{sp}}^*(\mathcal{X})=\mathcal{X}\mathbb{N}=X.
\end{equation}
Notice that, since the antisymmetric part of $\mathcal{X}\mathbb{N}$ vanishes, one has $\Bbb{N}^{-1}\mathcal{X}+\mathcal{X}^T\Bbb{N}^{-1}=0$, so that $\mathcal{X}\in\mathfrak{sp}(\R^{2k},\Bbb{N}^{-1})^*\simeq\mathfrak{sp}(\R^{2k},\Bbb{N}^{-1})$.  Then, the Bloch-Iserles Hamiltonian \eqref{BI_Ham} is given on $\mathfrak{sp}(\Bbb{R}^n,\Bbb{N}^{-1})^*$
by
\[
h(\mathcal{X})=\frac{1}{2}\operatorname{Tr}((\mathcal{X}\mathbb{N})^2)=\frac{1}{2}\operatorname{Tr}\left(\mathcal{X}(-\mathbb{N}^2)\mathcal{X}^T\right).
\]

\paragraph{Case 2: $\Bbb{N}$ of co-rank 1.} A more interesting situation occurs in Bloch-Iserles dynamics when $d=1$, that is  \makebox{$n=2k+1$} so that $\Bbb{N}$ is
of rank $2k$. In this case the Lie algebra isomorphism \eqref{psi}
becomes
\begin{equation}\label{LA_corank1}
\Psi:\left(\left(\operatorname{Sym}(\R^{2k})\,\circledS\,(\mathbb{R}^{2k})^*\right)\times_C\mathbb{R},[\cdot,\cdot]_C\right)\to\left(\operatorname{Sym}(\R^{2k+1}),\left[\cdot,\cdot\right]_\mathbb{N}\right),
\end{equation}
where the Lie bracket \eqref{BILieAlg} is now written as
\begin{equation}\label{Lie_bracket_General}
\left[(S_2,S_1,S_0),( S'_2, S'_1,
S'_0)\right]_C=\left(\left[S_2,S'_2\right]_{\bar{\mathbb{N}}}
,\,S_1\bar{\mathbb{N}} S'_2-
S'_1\bar{\mathbb{N}}S_{2\,},\,S_1\bar{\mathbb{N}}
S_1^{\prime\,T\,}\right).
\end{equation}
Notice that, for $d=1$, the space of matrices $\mathcal{M}_{d\times
2k}$ reduces to row $2k$-vectors, rather than column vectors in
$\R^{2k}$. This is the reason why we have inserted the notation
$(\R^{2k})^*$ for the space of linear forms to distinguish them from
ordinary vectors. Moreover, here we have chosen the suggestive
notation $(S,A,B)=(S_2,S_1,S_0)$ to show that in the corank $1$
case, \textit{we recover the Lie bracket underlying the dynamics of
the first three moments of the Vlasov system}, see
\eqref{mombrkt_symtens}. In terms of the variables $S_2,S_1,S_0$,
the Bloch-Iserles Lagrangian \eqref{BI_Lag_Psi} becomes
\[
\ell(S_2,S_1,S_0)=\frac{1}{2}\operatorname{Tr}(S_2^2)+|S_1|^2+2S_0^2.
\]
The expression \eqref{BI_Ham_Psi} for the Bloch-Iserles Hamiltonian
on ${\rm Sym}(\Bbb{R}^{2k})\times\Bbb{R}^{2k}\times\Bbb{R}$ becomes
\begin{equation}\label{BI_Ham_corank_1}
{\sf
h}(X_2,X_1,X_0)=\frac12\operatorname{Tr}(X_2^2)+\frac{1}{4}|X_1|^2+\frac{1}{8}X_0^2.
\end{equation}
\begin{remark}[The geodesic Vlasov equation underlying Bloch-Iserles dynamics]\normalfont
Notice that the above Hamiltonian $\mathsf{h}(X_0,X_1,X_2)$ can be written equivalently as a quadratic Vlasov Hamiltonian
\[
H(f)=\frac12\iint\! f({\bz})\mathcal{G}({\bz},{\bz}')f({\bz}')\,\de{\mathbf{z} }\,\de{\mathbf{z} }'
\,,\quad\text{ with }\quad
\mathcal{G}({\bf z,z}')=\frac18+\frac14{\bf z}\cdot{\bf z}'+\frac18({\bf z}\cdot{\bf z}')^2.
\]
Then, taking moments yields precisely the moment Hamiltonian $\mathsf{h}(X_0,X_1,X_2)$. Thus, again this means that the Bloch-Iserles system possesses a Vlasov formulation in terms of equation \eqref{geodesic-Vlasov}. As we shall see, this allows to identify momentum map solutions of the BI system that arise from the Klimontovich map \eqref{Klimontovich}. The problem of how the geometry of Vlasov dynamics is related to that of the BI system will be addressed in the second part of this paper.
\end{remark}

\begin{remark}[The role of the space $\mathcal{M}_{d\times 2k}$]\normalfont Consider the general case when $\mathbb{N}$ has corank $d$. One can use the isomorphism \eqref{Lie_algebra_isom} (upon
replacing $\Bbb{N}$ by $\bar{\Bbb{N}}$) to identify the Lie algebra $\left(\operatorname{Sym}(\mathbb{R}^{2k})\,\circledS\,\mathcal{M}_{d\times
2k}\right)\times\operatorname{Sym}(\mathbb{R}^d)$ with
\[
\left(\mathfrak{sp}(\Bbb{R}^{2k},\bar{\mathbb{N}}^{-1})\,\circledS\,\mathcal{M}_{d\times
2k}\right)\times_C\operatorname{Sym}(\Bbb{R}^{d}).
\]
If $d=1$ we thus get the Lie algebra
\[
\left(\mathfrak{sp}(\Bbb{R}^{2k},\bar{\mathbb{N}}^{-1})\,\circledS\,(\mathbb{R}^{2k})^*\right)\times_C\mathbb{R}.
\]
However, as we shall see in the next section, the case $d=1$ suggests the need of incorporating ordinary vectors in
$\R^{2k}$ (rather than linear forms in
$(\R^{2k})^*$) within the Lie algebra structure. We shall show how to slightly modify the
isomorphism
\[
\operatorname{Sym}(\R^{2k})\,\circledS\,(\R^{2k})^*\to\mathfrak{sp}(\R^{2k},\bar{\Bbb{N}}^{-1})\,\circledS\,(\R^{2k})^*,\quad (S_2,S_1)\mapsto (\Bbb{N}S_2,S_1)
\]
into an isomorphism
$\operatorname{Sym}(\R^{2k})\,\circledS\,(\R^{2k})^*\to\mathfrak{sp}(\R^{2k},\bar{\Bbb{N}}^{-1})\,\circledS\,\R^{2k}$,
which naturally leads to the Lie algebra of the Jacobi group
$\operatorname{Jac}(\R^{2k},\bar{\Bbb{N}}^{-1})=\operatorname{Sp}(\R^{2k},\bar{\Bbb{N}}^{-1})\,\circledS\,\operatorname{H}(\R^{2k},\bar{\Bbb{N}}^{-1})$.
Consequently, in the general corank $d$ case, we shall need an isomorphism that
also affects the space $\mathcal{M}_{d\times 2k}$.
\end{remark}

\subsection{Summary of main results}
The results in this paper emerge from the discovery that the integrable Bloch-Iserles system possesses a Vlasov description in terms of statistical moments \cite{HoTr2009}. Then, this paper uses the polynomial algebra $\operatorname{Pol}(\Bbb{R}^{2k})\subset\mathcal{F}(\Bbb{R}^{2k})$ associated to the Vlasov equation in order to identify the Jacobi group as the Lie symmetry underlying the integrable Bloch-Iserles (BI) system. We  show that the general case when $\operatorname{corank}(\Bbb{N})>1$ requires a suitable generalization of the Jacobi group that is different from other generalizations already appeared in the literature. The Lie-Poisson structure of the BI system is then analyzed in detail and we show how this structure leads to a natural momentum map, which in turn arises from the Vlasov single-particle solution (Klimontovich).

 The relation between Vlasov dynamics and the Bloch-Iserles system naturally leads to the question of how the Jacobi group is a subgroup of the strict contact transformations, which underlie Vlasov dynamics. This question is addressed in the second part of this paper, where explicit subgroup inclusions are presented.  Also, we show how the generalized Jacobi group is a natural generalization of $\operatorname{Cont}(\Bbb{R}^{2k+1}, \theta )$.  This particular result arises as a consequence of some recent developments on Abelian extensions of the quantomorphism group \cite{Vi2010,NeVi2003}. After the relation between the BI system and its underlying Vlasov description is completely characterized, the momentum maps accompanying the Klimontovich solution are analyzed in detail.

\section{The symmetry group of the Bloch-Iserles system}
\label{symmetry_group_BI}

As recalled above, when $\mathbb{N}$ is invertible the Lie algebra
$\left({\rm
Sym}(\R^{2k}),\left[\cdot,\,\cdot\right]_\mathbb{N}\right)$ is
isomorphic to the Lie algebra of the symplectic group
$\operatorname{Sp}(\mathbb{R}^{2k},\mathbb{N}^{-1})$. Therefore, the
associated Bloch-Iserles system is the convective representation of
the geodesics on
$\operatorname{Sp}(\mathbb{R}^{2k},\mathbb{N}^{-1})$ relative to a
left invariant Riemannian metric. More precisely, a curve $g(t)\in
\operatorname{Sp}(\mathbb{R}^{2k},\mathbb{N}^{-1})$ is a geodesic if
and only if $X(t)=g(t)^{-1}\dot g(t)$ solves the Bloch-Iserles
system.

In this section we shall show that a similar geodesic interpretation exists for any antisymmetric matrix $\mathbb{N}$. In order to do this, we will exhibit a Lie group whose Lie algebra is given by $\left({\rm
Sym}(\R^{n}),\left[\cdot,\,\cdot\right]_\mathbb{N}\right)$, where $\mathbb{N}$ is an arbitrary antisymmetric matrix of corank $d$. We first treat the case $d=1$.

\subsection{Bloch-Iserles dynamics on the Jacobi group}

In order to clarify the occurrence of the symplectic structure $\mathbb{J}$ and its Poisson matrix operator $\mathbb{N}=-\mathbb{J}^{-1}$ (see remark \ref{remark-on-N}) as well as the role of the symplectic space $\mathbb{R}^{2k}$, it is easier to use an intrinsic formulation rather than the matrix notation. We thus consider an arbitrary finite dimensional symplectic vector space $(V,\Omega)$ with  symplectic form $\Omega$. The corresponding notation is introduced in Appendix \ref{notation}, to which the reader is addressed for further details.

\subsubsection{The Heisenberg and Jacobi groups}

 The
{\it Heisenberg group} associated to $(V,\Omega)$ is the central
extension of the symplectic vector space $V$ (viewed as an Abelian group),
relative to the group two cocycle $\mathscr{B}:V\times V\to\Bbb{R}$ 
given by
\begin{equation}\label{groupcocycle}
\mathscr{B}(u,v):=\frac{1}{2}\Omega(u,v),
\end{equation}
so that the group multiplication is:
\[
(u,a)(v,b)=\left(u+v,a+b+\frac{1}{2}\Omega(u,v)\right).
\]
On the other hand, the Lie algebra is given by the central extension
$\mathfrak{h}(V,\Omega)=V\times_{\mathscr{B}}\mathbb{R}$ of the
vector space $V$, relative to the Lie algebra two cocycle $\mathscr{B}$. The Lie bracket on $\mathfrak{h}(V,\Bbb{R})$ is
\[
[(u,a),(v,b)]=(0,\Omega(u,v)).
\]
Notice that the (\textit{left}) Lie-Poisson bracket on the dual Lie algebra $\mathfrak{h}^*(V,\Omega)=V^*\times\mathbb{R}$ has the particular form
\begin{equation}\label{LP_Heisenberg}
\{f,g\}_-(k,m)=-m\,\Omega\left(\frac{\delta f}{\delta
k},\frac{\delta g}{\delta k}\right)=-m\left\{f,g\right\}_{V^*},
\end{equation}
where $\{\,,\}_{V^*}$ denotes the symplectic Poisson bracket induced on $V^*$. This bracket is related to the Poisson bracket on $V$ by the Poisson tensor $\Pi^\sharp:V^*\rightarrow V$.
For $m\in\mathbb{R}$ fixed, the hyperplanes $V^*\times\{m\}$ are Poisson submanifolds of $\mathfrak{h}^*(V,\Omega)$.

\medskip

There is an action of the symplectic group
$\operatorname{Sp}(V,\Omega)$ on $\operatorname{H}(V,\Omega)$ given
by the representation on the first factor, that is
\[
(u,a)\mapsto g(u,a):=(gu,a),\quad g\in\operatorname{Sp}(V,\Omega).
\]
It is readily seen that $\operatorname{Sp}(V,\Omega)$ acts by group homomorphisms. Therefore, we can form the associated semidirect product
\[
\operatorname{Jac}(V,\Omega):=\operatorname{Sp}(V,\Omega)\,\circledS\,\operatorname{H}(V,\Omega),
\]
called the \textit{Jacobi group}. The group multiplication is given
by
\[
(g,u,a)(h,v,b)=\left(gh,u+gv,a+b+\frac{1}{2}\Omega(u,gv)\right),
\]
and the associated bracket on the Lie algebra
$\mathfrak{jac}(V,\Omega)=\mathfrak{sp}(V,\Omega)\,\circledS\,\mathfrak{h}(V,\Omega)$
reads
\begin{equation}\label{Lie_bracket_jac}
\left[(\mathcal{S}_2,\mathcal{S}_1,\mathcal{S}_0),(\mathcal{S}'_2,\mathcal{S}'_1,\mathcal{S}'_0)\right]=\left(\left[\mathcal{S}_2,\mathcal{S}'_2\right],\mathcal{S}_{2\,}\mathcal{S}'_1-\mathcal{S}'_{2\,}\mathcal{S}_{1\,},\Omega(\mathcal{S}_1,\mathcal{S}'_1)\right).
\end{equation}
In the special case $V=\Bbb{R}^2$, coadjoint motion on $\operatorname{Jac}(\Bbb{R}^2)$ has been studied in \cite{Berndt2003,Berndt2006,Yang2003}. As we shall see in the next section, the Jacobi group (and its Lie algebra, above) fully characterizes the Bloch-Iserles system, in the case $\operatorname{corank}{{\Bbb{N}}}=1$.

\subsubsection{Geodesic flows on the Jacobi group} From the
general setting above, we may now characterize the case $\operatorname{corank}\Bbb{N}=1$ of
Bloch-Iserles dynamics in a completely intrinsic fashion, that
clarifies the occurrence of the symplectic structure $\mathbb{J}=-\mathbb{N}^{-1}$ (see remark \eqref{remark-on-N}) as well as the the role of the
symplectic space $\mathbb{R}^{2k}$ and its dual. Indeed, we
recognize that the intrinsic formulation of the Lie algebra
\[
\left(\operatorname{Sym}(\R^{2k})\,\circledS\,(\mathbb{R}^{2k})^*\right)\times_C\mathbb{R}\
\ni\,  (S,A,B)
\]
 is given by
\[
\left(\operatorname{Sym}(V)\,\circledS\,
V^*\right)\times_C\mathbb{R}\ \ni\, (\sigma,\alpha,\beta),
\]
where
\[
\operatorname{Sym}(V)=\left\{\sigma:V\times V\to\R \
\vert\ \sigma(u,v)=\sigma(v,u),\,\ \forall\,u,v\in V\right\}
\]
is the space of symmetric bilinear forms on $V$. The Lie bracket $[\,,]_{\mathbb{N}}$ is intrinsically written as
\[
[\sigma,\sigma']_\Pi=\sigma\circ \Pi^\sharp\circ \sigma'-\sigma'\circ
\Pi^\sharp\circ \sigma,
\]
where we canonically identify an arbitrary symmetric bilinear form
$\sigma:V\times V\rightarrow\mathbb{R}$ with a symmetric linear map
$\sigma^\flat:V\rightarrow V^*$, via the usual relation
\[
\left\langle\sigma^\flat(u),v\right\rangle:=\sigma(u,v).
\]
For simplicity, we shall suppress the suffix $\flat$ since the
interpretation of $\sigma$ will be clear from the context.

The semidirect product $\operatorname{Sym}(V)\,\circledS\,
V^*$ is taken
with respect to the Lie algebra action
\[
\alpha\mapsto\sigma\cdot \alpha=-\alpha\circ \Pi^\sharp\circ\sigma
\]
and the Lie algebra $2$-cocycle $C$ is given by the Poisson tensor:
\[
C((\sigma,\alpha),(\sigma',\alpha'))=\Pi(\alpha,\alpha').
\]
Thus, the Lie bracket on $\left(\operatorname{Sym}(V)\,\circledS\,
V^*\right)\times_{C}\mathbb{R}$ reads
\begin{equation}\label{Lie_bracket_corank_one}
\left[(\sigma,\alpha,\beta),({\sigma}',{\alpha}',{\beta}')\right]=\left(\sigma\circ
\Pi^\sharp\circ {\sigma}'-{\sigma}'\circ \Pi^\sharp\circ
\sigma,\,\alpha\circ \Pi^\sharp\circ{\sigma}'-{\alpha}'\circ
\Pi^\sharp\circ\sigma,\,\Pi(\alpha,{\alpha}')\right).
\end{equation}
One can check that when $V=\mathbb{R}^{2k}$ and the matrix notation $[\Pi]=\bar{\mathbb{N}}$ is used, we recover the Lie algebra action \eqref{LieAlgAct}, the cocycle \eqref{cocycle1}, and the Lie bracket \eqref{Lie_bracket_General} given in the previous section for the case $d=1$.

\medskip

The following lemma is a crucial result. It says that, for the corank one case, the Lie algebra underlying the Bloch-Iserles is isomorphic to the Lie algebra of the Jacobi group.

\begin{lemma}\label{lemma:LAisomorphism}
With the notation above, the map
\[
\Gamma:\left(\operatorname{Sym}(V)\,\circledS\,
V^*\right)\times_{C}\mathbb{R}\rightarrow\mathfrak{jac}(V,\Omega),\quad(\sigma,\alpha,\beta)\mapsto
(\mathcal{S}_2,\mathcal{S}_1,\mathcal{S}_0)=(\Pi^\sharp\circ\sigma,\Pi^\sharp(\alpha),\beta).
\]
is a Lie algebra isomorphism.
\end{lemma}
\paragraph{Proof.} The linear map $\Gamma$ is clearly an isomorphism. We now show that it respects the Lie brackets. Applying $\Gamma$ to the Lie bracket \eqref{Lie_bracket_corank_one} we get the expression
\begin{equation}\label{auxialiary-expression}
\left(\Pi^\sharp\circ\sigma\circ
\Pi^\sharp\circ {\sigma}'-\Pi^\sharp\circ{\sigma}'\circ \Pi^\sharp\circ
\sigma,\,\Pi^\sharp(\alpha\circ \Pi^\sharp\circ{\sigma}')-\Pi^\sharp({\alpha}'\circ
\Pi^\sharp\circ\sigma),\,\Pi(\alpha,{\alpha}')\right).
\end{equation}
Using the formulas
\[
\alpha\circ \Pi^\sharp\circ\sigma=-\sigma\left(\Pi^\sharp(\alpha)\right)\quad
\text{and}\quad
\Omega\left(\Pi^\sharp(\alpha),\Pi^\sharp({\alpha}')\right)=\Pi(\alpha,{\alpha}'),
\]
the expression \eqref{auxialiary-expression} can be rewritten as
\[
\left(\Pi^\sharp\circ\sigma\circ \Pi^\sharp\circ
{\sigma}'-\Pi^\sharp\circ{\sigma}'\circ \Pi^\sharp\circ
\sigma,\,(\Pi^\sharp\circ\sigma)(\Pi^\sharp(\alpha'))-(\Pi^\sharp\circ\sigma')(\Pi^\sharp(\alpha)),\,\Omega\!\left(\Pi^\sharp(\alpha),\Pi^\sharp({\alpha}')\right)\right).
\]
This recovers the Lie bracket \eqref{Lie_bracket_jac} of the Jacobi
Lie algebra, evaluated on the elements $\Gamma(\sigma,\alpha,\beta)$,
$\Gamma(\sigma',\alpha',\beta')$ of
$\mathfrak{jac}(V,\Omega).\qquad\blacksquare$

\medskip

At this point, upon fixing
\[
(V,\Omega)=(\R^{2k},-\bar{\Bbb{N}}^{-1}),
\]
the Lie bracket on $\mathfrak{jac}(\R^{2k},-\bar{\Bbb{N}}^{-1})$
yields the Lie algebra structure \eqref{Lie_bracket_General} for the
Bloch-Iserles system \eqref{BI} in the case
$\operatorname{corank}(\Bbb{N})=1$. In this case, the above
isomorphism is given by
\begin{equation}\label{Gamma_corankone}
\Gamma:\left(\operatorname{Sym}(\R^{2k})\,\circledS\,
(\mathbb{R}^{2k})^*\right)\times_C\mathbb{R}\rightarrow
\mathfrak{jac}(\R^{2k},-\bar{\mathbb{N}}^{-1}),
\end{equation}
\[
(S_2,S_1,S_0)\mapsto
\left(\mathcal{S}_2,\mathcal{S}_1,\mathcal{S}_0\right)=\left(\bar{\mathbb{N}}S_2,\bar{\mathbb{N}}S_1^T,S_0\right)
.
\]
which produces the Bloch-Iserles dynamics on
$\mathfrak{jac}(\R^{2k},-\bar{\mathbb{N}}^{-1})$.

\begin{remark}[Nature of the Lie algebra isomorphism $\Gamma$]\normalfont
At this point, we observe how the isomorphism
$\Gamma:\left(\operatorname{Sym}(V)\,\circledS\,
V^*\right)\times_{\!C}\mathbb{R}\rightarrow\mathfrak{jac}(V,\Omega)$,
affects both the $\operatorname{Sym}(V)$ and the $V^*$ components in
the Lie algebra $\left(\operatorname{Sym}(V)\,\circledS\,
V^*\right)\times_{\!C}\mathbb{R}$. Indeed, while the map
\[
\operatorname{Sym}(V)\ni\sigma\mapsto \Pi^\sharp\circ
\sigma\in\mathfrak{sp}(V,\Omega)
\] 
was already introduced in
\eqref{Lie_algebra_isom}, the map 
\[
V^*\ni\alpha\mapsto
\Pi^\sharp\circ \alpha\in V
\]
is needed here too. This fact is crucial
for understanding the geometric construction underlying the
Bloch-Iserles system and it is necessary in order to match the
correct Lie algebra
$\mathfrak{jac}(V,\Omega)=\mathfrak{sp}(V,\Omega)\,\circledS\,\mathfrak{h}(V,\Omega)$.
\end{remark}

\medskip

We can now rewrite the Bloch-Iserles Lagrangian on
$\mathfrak{jac}(\R^{2k},-\bar{\mathbb{N}}^{-1})$. We find
\[
\ell(\mathcal{S}_2,\mathcal{S}_1,\mathcal{S}_0)=\frac{1}{2}\operatorname{Tr}(\mathcal{S}_2^T(-\bar{\mathbb{N}}^{-2})\mathcal{S}_2)+\mathcal{S}_1^T(-\bar{\mathbb{N}}^{-2})\mathcal{S}_1+2\mathcal{S}_0^2.
\]
The dual isomorphism arises from the Killing form pairing
$\left\langle\mathcal{X,S}\right\rangle=\operatorname{Tr}(\mathcal{XS})$
on $\mathfrak{sp}(\Bbb{R}^{2k},-\Bbb{N}^{-1})$ and it is given by
\[
\mathfrak{jac}(\R^{2k},-\bar{\mathbb{N}}^{-1})^*\rightarrow
\operatorname{Sym}(\R^{2k})\times\mathbb{R}^{2k}\times\mathbb{R}
\]
\begin{equation}\label{isom1}
(\mathcal{X}_2,\mathcal{X}_1,\mathcal{X}_0)\mapsto
(X_2,X_1,X_0)=\left(\mathcal{X}_2\bar{\mathbb{N}},-\bar{\mathbb{N}}\mathcal{X}_1^T,\mathcal{X}_0\right).
\end{equation}
This allows us to write the Bloch-Iserles Hamiltonian on the dual of the Lie algebra of the Jacobi group as
\begin{align*}
h(\mathcal{X}_2,\mathcal{X}_1,\mathcal{X}_0)&=\frac12\operatorname{Tr}(X_2^2)+\frac{1}{4}|X_1|^2+\frac{1}{8}X_0^2\\
&=\frac{1}{2}\operatorname{Tr}\left(\mathcal{X}_2(-\bar{\mathbb{N}}^2)\mathcal{X}_2^T\right)+\frac{1}{4}\mathcal{X}_1(-\bar{\mathbb{N}}^2)\mathcal{X}_1^T+\frac{1}{8}\mathcal{X}_0^2,
\end{align*}
where we used the expression of the Hamiltonian \eqref{BI_Ham_corank_1}.

Using the Euler-Poincar\'e reduction theorem and the results obtained above, we obtain the following theorem.

\begin{theorem} Assume that the antisymmetric matrix $\mathbb{N}$ has $\operatorname{corank}(\Bbb{N})=1$. Then the Bloch-Iserles system $\dot X(t)=[X(t)^2,\mathbb{N}]$ describes geodesic motion on the Jacobi group $\operatorname{Jac}(2k,-\bar{\mathbb{N}}^{-1})$ relative to the left invariant metric given at the identity by the Lagrangian
\[
\ell( \mathcal{S} _0 , \mathcal{S} _1 , \mathcal{S} _2 )=\frac{1}{2}\operatorname{Tr}(\mathcal{S}_2^T(-\bar{\mathbb{N}}^{-2})\mathcal{S}_2)+\mathcal{S}_1^T(-\bar{\mathbb{N}}^{-2})\mathcal{S}_1+2\mathcal{S}_0^2.
\]
\end{theorem}
\noindent\textbf{Proof.} We know that, in the general case, the Bloch-Iserles system is an Euler-Poincar\'e equation on the Lie algebra $\operatorname{Sym}(\mathbb{R}^n,[\,,]_\mathbb{N})$. When $\mathbb{N}$ has corank one, this Lie algebra is isomorphic to the Jacobi Lie algebra, by the preceding Lemma. The result follows by applying the Euler-Poincar\'e reduction to the Jacobi group $\operatorname{Jac}(\mathbb{R}^{2k},-\bar{\mathbb{N}}^{-1})$ relative to the left invariant Riemannian metric induced by the Bloch-Iserles Lagrangian. $\qquad\blacksquare$

\subsection{Bloch-Iserles dynamics on the generalized Jacobi group}

Once we have characterized the special case $\operatorname{corank}(\Bbb{N})=1)$, we are now ready to identify the Lie group symmetry of Bloch-Iserles dynamics for the most general case $\operatorname{corank}(\Bbb{N})>1$. This is the subject of the present section.

\subsubsection{Intrinsic definition of the Bloch-Iserles
Lie algebra} As we have seen, if $\mathbb{N}$ has corank $d$, then
we split $\mathbb{R}^n$ in such a way that the matrix representation
of $\mathbb{N}$ has the form
\[
\mathbb{N}=\left[
\begin{array}{cc}
\bar{\Bbb{N}} & 0
\\
0 & 0
\end{array}
\right],
\]
where $\bar{\Bbb{N}}$ is a $2k\times 2k$ invertible matrix, with
$n=2k+d$. In general, if $V'$ is a vector space endowed with a
antisymmetric (degenerate) bilinear form $\Omega'$, then we can write
$V'=V\oplus W$, where $W=\operatorname{ker}(\Omega')$ and $V$ carries the  symplectic form $\Omega:=\Omega'|_V$. As in \eqref{psi}, we have an
isomorphism
\[
\Psi:\left(\left(\operatorname{Sym}(V)\,\circledS\,L(V,W^*)\right)\times_C\operatorname{Sym}(W),[\cdot,\cdot]_C\right)\rightarrow\left(\operatorname{Sym}(V\oplus
W),[\cdot,\cdot]_\mathbb{N}\right)
\]
given by
\[
\Psi(\sigma,\alpha,\beta):=\left[
\begin{array}{cc}
\sigma & \alpha^*
\\
\alpha & 2\beta
\end{array}
\right],
\]
where we use the matrix notation to write a symmetric bilinear form
on $V'$ relative to the decomposition $V'=V\oplus W$. Here, $L(V,W^*)$ denotes linear maps $V\to W^*$ as usual. The semidirect
product $\operatorname{Sym}(V)\,\circledS\,L(V,W^*)$ involves the Lie algebra action of $\operatorname{Sym}(V)$
on $L(V,W^*)$ given by
\[
\alpha\mapsto \sigma\cdot\alpha=-\alpha\circ \Pi^\sharp\circ \sigma\
\quad\text{ with }\quad  \sigma\in\operatorname{Sym}(V),\,\alpha\in
L(V,W^*)
\]
while the central extension by $\operatorname{Sym}(W)$ is given by
the cocycle
\[
C\left((\sigma,\alpha),(\sigma',\alpha')\right)=\Pi(\alpha^*(\cdot),\alpha^{\prime\,*}(\cdot))^{\rm sym\,}\in\operatorname{Sym}(W),
\]
where $\Pi(\alpha^*(\cdot),\alpha^{\prime\,*}(\cdot))^{\rm sym\,}$
denotes the symmetric part of the bilinear map
\[
(w,w')\mapsto \Pi(\alpha^*(w),\alpha^{\prime\,*}(w'))\,,
\quad
w,w'\in W.
\]
The Lie bracket on the left hand side is readily seen to be
\begin{equation}\label{BILieAlg_abstract}
\left[(\sigma,\alpha,\beta),(\sigma',\alpha',\beta')\right]_C=\left([\sigma,\sigma']_\Pi,\,\alpha\circ \Pi^\sharp\circ\sigma'-\alpha'\circ \Pi^\sharp\circ \sigma\,,
\Pi(\alpha^*(\cdot),\alpha^{\prime\,*}(\cdot))^{\rm sym\,}\right),
\end{equation}
where
\[
[\sigma,\sigma']_\Pi=\sigma\circ \Pi^\sharp\circ \sigma'-\sigma'\circ
\Pi^\sharp\circ \sigma.
\]

The goal is now to obtain the generalization of Lemma
\ref{lemma:LAisomorphism} to the case
$\operatorname{corank}(\Bbb{N})=d$. We first define a generalization
of the Heisenberg and Jacobi groups.

\subsubsection{The generalized Heisenberg group}
This generalization is given by the central extension
\[
\operatorname{H}(V,\Omega;W)=L(W,V)\times_\mathscr{B}\operatorname{Sym}(W)\ni(\mathcal{A},\mathcal{B}),
\]
of the space of linear maps $L(W,V)$ from $W$ to $V$ relative to the $\operatorname{Sym}(W)$-valued group 2-cocycle:
\[
\mathscr{B}:L(W,V)\times L(W,V)\to\operatorname{Sym}(W),\qquad
\mathscr{B}(\mathcal{A},\mathcal{A}'):=\frac{1}{2}\Omega(\mathcal{A}(\cdot),\mathcal{A}'(\cdot))^{\sym}
\]
so that
\[
\mathscr{B}(\mathcal{A},\mathcal{A}')(w_1,w_2)=
\frac14\Big(\,\Omega\!\left(\mathcal{A}(w_1),\mathcal{A}'(w_2)\right)
+\Omega\!\left(\mathcal{A}(w_2),\mathcal{A}'(w_1)\right)\Big).
\]
The group multiplication is thus given by
\begin{equation}\label{gen_Heis_multiplication}
(\mathcal{A},\mathcal{B})(\mathcal{A}',\mathcal{B}')=(\mathcal{A}+\mathcal{A}',\mathcal{B}+\mathcal{B}'+\mathscr{B}(\mathcal{A},\mathcal{A}'))
\end{equation}
and the Lie bracket is
\[
[(\mathcal{A},\mathcal{B}),(\mathcal{A}',\mathcal{B}')]=(0,2\mathscr{B}(\mathcal{A},\mathcal{A}'))=\left(0,\Omega(\mathcal{A}(\cdot),\mathcal{A}'(\cdot))^{\sym}\right).
\]
If $d=\operatorname{dim}(W)=1$ we recover the Heisenberg group since
in this case $L(W,V)=V$ and
\[
\mathscr{B}(\mathcal{A},\mathcal{A}')=\frac{1}{2}\Omega(\mathcal{A}(\cdot),\mathcal{A}'(\cdot))^{\sym}=\frac{1}{2}\Omega(\mathcal{A},\mathcal{A}').
\]
Notice that other generalizations of the Heisenberg group are also available; see e.g. \cite{Yang2002}. However, to the authors' knowledge, the Lie group presented above appears in the literature for the first time.
\begin{remark}[Lie-Poisson bracket on $\mathfrak{h}^*(V,\Omega;W)$]\normalfont
Upon choosing $(V,\Omega)=(\mathbb{R}^{2k},\mathbb{J})$ and
$W=\mathbb{R}^d$, we have $L(W,V)=\mathcal{M}_{2k\times d}$ and the
Lie bracket reads
\[
[(\mathcal{A},\mathcal{B}),(\mathcal{A}',\mathcal{B}')]=\left(0,(\mathcal{A}^T\mathbb{J}\mathcal{A}')^{sym}\right).
\]
We thus obtain the (\textit{left}) Lie-Poisson bracket
\[
\{f,g\}_-(\mathcal{K},\mathcal{M})=-\operatorname{Tr}\left(\mathcal{M}\left(\frac{\partial f}{\partial\mathcal{K}}\right)^T\mathbb{J}\,\frac{\partial g}{\partial\mathcal{K}}\right),\quad (\mathcal{K},\mathcal{M})\in\mathcal{M}_{d\times 2k}\times\operatorname{Sym}(\mathbb{R}^d).
\]
This is the generalization of the bracket \eqref{LP_Heisenberg} to the case where $d\geq 1$.
\end{remark}

\subsubsection{The generalized Jacobi group}\label{def_gen_Jac}
We now let the symplectic group $\operatorname{Sp}(V,\Omega)$ act on
the generalized Heisenberg group $\operatorname{H}(V,\Omega;W)$ from
the left by the action
\[
(\mathcal{A},\mathcal{B})\mapsto (g\circ\mathcal{A},\mathcal{B}).
\]
Using the expression \eqref{gen_Heis_multiplication} for the multiplication in the general Heisenberg group
and the property
$\mathscr{B}(g\mathcal{A},g\mathcal{A}')=\mathscr{B}(\mathcal{A},\mathcal{A}')$, for all $g\in \operatorname{Sp}(V,\Omega)$,
it follows that we have an action by homomorphisms.
 Thus, we can
define the group semidirect product
\[
\operatorname{Jac}(V,\Omega;W):=\operatorname{Sp}(V,\Omega)\,\circledS\,\operatorname{H}(V,\Omega;W)
\,,
\]
which we call the \textit{generalized Jacobi group}. We now compute the associated Lie bracket. In general, for a semidirect product
$G\,\circledS\,H$ of two groups we have the formula
\[
[(\xi,\eta),(\xi',\eta')]=([\xi,\xi'],\xi\cdot\eta'-\xi'\cdot\eta+[\eta,\eta']).
\]
In our case, the first factor is $[\mathcal{S},\mathcal{S}']$
and the second factor is
\[
\mathcal{S}\cdot
(\mathcal{A}',\mathcal{B}')-\mathcal{S}'\cdot
(\mathcal{A},\mathcal{B})+(0,2\mathscr{B}(\mathcal{A},\mathcal{A}'))=\left(\mathcal{S}\circ\mathcal{A}'-\mathcal{S}'\circ\mathcal{A},\Omega(\mathcal{A}(\cdot),\mathcal{A}'(\cdot))^{sym}\right),
\]
we thus obtain the Lie bracket associated to the Lie algebra $\mathfrak{jac}(V,\Omega;W)$ of the generalized Jacobi group:
\begin{equation}\label{Lie_bracket_Jac}
[(\mathcal{S},\mathcal{A},\mathcal{B}),(\mathcal{S}',\mathcal{A}',\mathcal{B}')]=\left(
[\mathcal{S},\mathcal{S}'],\mathcal{S}\circ\mathcal{A}'-\mathcal{S}'\circ\mathcal{A},\Omega(\mathcal{A}(\cdot),\mathcal{A}'(\cdot))^{sym}\right).
\end{equation}

\subsubsection{Geodesic flows on the generalized Jacobi group}
We now formulate the generalization of Lemma
\ref{lemma:LAisomorphism} to the case
$\operatorname{corank}(\Bbb{N})=d$.

\begin{lemma}\label{lemma:LAisomorphism_d}
With the notation above, the map
\[
\Gamma:\operatorname{Sym}(V\oplus W)\simeq\left(\operatorname{Sym}(V)\,\circledS\,
L(V,W^*)\right)\times\operatorname{Sym}(W)\rightarrow\mathfrak{jac}(V,\Omega;W),
\]
\[
(\sigma,\alpha,\beta)\mapsto
(\mathcal{S},\mathcal{A},\mathcal{B})=(B^\sharp\circ\sigma,B^\sharp\circ\alpha^*,\beta).
\]
is a Lie algebra isomorphism.
\end{lemma}
\paragraph{Proof.} We apply the imomorphism $\Gamma$ to the Lie bracket \eqref{BILieAlg_abstract}. For the first component we obtain the desired expression $B^\sharp\circ\sigma\circ B^\sharp\circ \sigma'-B^\sharp\circ\sigma'\circ B^\sharp\circ \sigma=[\mathcal{S},\mathcal{S}']$. For the second component, we get
\begin{align*}
B^\sharp\circ (\alpha\circ B^\sharp\circ\sigma')^*-B^\sharp\circ (\alpha'\circ B^\sharp\circ\sigma)^*&=-B^\sharp\circ \sigma'\circ B^\sharp\circ \alpha^*+B^\sharp\circ \sigma\circ B^\sharp\circ \alpha^{\prime \,*}\\
&=\mathcal{S}\circ\mathcal{A}'-\mathcal{S}'\circ\mathcal{A},
\end{align*}
which is exactly the second component of the Lie bracket on $\mathfrak{jac}(V,\Omega;W)$. For the third component, we have
\[
B(\alpha^*(\cdot),\alpha^{\prime\,*}(\cdot))^{\rm sym\,}=\Omega(B^\sharp(\alpha^*(\cdot)),B^\sharp(\alpha^{\prime\,*}(\cdot)))^{\rm sym\,}=\Omega(\mathcal{A}(\cdot),\mathcal{A}'(\cdot))^{\rm sym\,}
\]
as required.$\qquad\blacksquare$

\medskip

\noindent
Then, upon fixing
\[
(V,\Omega)=(\R^{2k},-\bar{\Bbb{N}}^{-1})\quad\text{and}\quad
W=\mathbb{R}^d,
\]
we have
\[
\operatorname{H}(\mathbb{R}^{2k},-\bar{\mathbb{N}}^{-1};\mathbb{R}^d)=\mathcal{M}_{2k
\times d}\times_\mathscr{B}\operatorname{Sym}(\mathbb{R}^d)\ni
(\mathcal{A},\mathcal{B})
\]
and the cocycle reads
$\mathscr{B}(\mathcal{A},\mathcal{A}')=-\frac{1}{2}(\mathcal{A}^T\mathbb{N}^{-1}\mathcal{A}')^{\rm sym}$.
Then, the Lie bracket on the generalized Jacobi Lie algebra
$\mathfrak{jac}(\R^{2k},-\bar{\Bbb{N}}^{-1};\R^d)$ yields the Lie
algebra structure \eqref{BILieAlg} for the Bloch-Iserles system
\eqref{BI} in the case $\operatorname{corank}(\Bbb{N})=d$. In this
case, the above isomorphism is given by
\[
\Gamma: \left(\operatorname{Sym}(\R^{2k})\,\circledS\, \mathcal{M}_{d\times
2k}\right)\times_C\operatorname{Sym}(\mathbb{R}^d)\rightarrow
\mathfrak{jac}(\R^{2k},-\bar{\mathbb{N}}^{-1};\mathbb{R}^d),
\]
\[
(S,A,B)\mapsto
\left(\mathcal{S},\mathcal{A},\mathcal{B}\right)=\left(\bar{\mathbb{N}}S,\bar{\mathbb{N}}A^T,B\right)
.
\]
 
\begin{remark}[Specializations of the isomorphism $\Gamma$]\label{SpecialGamma}\normalfont Notice that, while the first component of $\Gamma$ coincides with the isomorphism \eqref{Lie_algebra_isom}, another Lie algebra isomorphism 
\[
\Gamma_\mathfrak{h}:\mathcal{M}_{d\times
2k}\times_C\operatorname{Sym}(\mathbb{R}^d)\to\mathfrak{h}(\R^{2k},-\bar{\mathbb{N}}^{-1};\mathbb{R}^d)
\]
arises from the last two components of $\Gamma$.
\end{remark}

With the help of the isomorphism $\Gamma$, we can now rewrite the
Bloch-Iserles Lagrangian on the Lie algebra
$\mathfrak{jac}(\R^{2k},-\bar{\mathbb{N}}^{-1};\mathbb{R}^d)$. We
find
\[
\ell(\mathcal{S},\mathcal{A},\mathcal{B})=\frac{1}{2}\operatorname{Tr}(\mathcal{S}^T(-\bar{\mathbb{N}}^{-2})\mathcal{S})+\operatorname{Tr}\left(\mathcal{A}^T(-\bar{\mathbb{N}}^{-2})\mathcal{A}\right)+2\operatorname{Tr}(\mathcal{B}^2).
\]
The dual isomorphism is given by
\begin{equation}\label{gamma_star}
\mathfrak{jac}(\R^{2k},-\bar{\mathbb{N}}^{-1};\mathbb{R}^d)^*\rightarrow
\operatorname{Sym}(2k)\times\mathcal{M}_{2k\times
d}\times\operatorname{Sym}(d)
\end{equation}
\[
(\mathcal{X},\mathcal{K},\mathcal{M})\mapsto
(X,K,M)=\left(\mathcal{X}\bar{\mathbb{N}},-\bar{\mathbb{N}}\mathcal{K}^T,\mathcal{M}\right),
\]
where $(\mathcal{X},\mathcal{K},\mathcal{M})\in \operatorname{Sp}(\mathbb{R}^{2k},-\bar{\mathbb{N}}^{-1})\times\mathcal{M}_{d\times 2k}\times\operatorname{Sym}(\mathbb{R}^d)$.
This allows us to write the Bloch-Iserles Hamiltonian on the dual of
the Lie algebra of the Lie group
$\operatorname{Jac}(2k,d,-\bar{\mathbb{N}}^{-1})$ as
\begin{equation}\label{BI_Hamiltonian_on_Jac}
h(\mathcal{X},\mathcal{K},\mathcal{M})=\frac{1}{2}\operatorname{Tr}\left(\mathcal{X}(-\bar{\mathbb{N}}^2)\mathcal{X}^T\right)+\frac{1}{4}\operatorname{Tr}(\mathcal{K}(-\bar{\mathbb{N}}^2)\mathcal{K}^T)+\frac{1}{8}\operatorname{Tr}(\mathcal{M}^2).
\end{equation}
Upon using the Euler-Poincar\'e reduction theorem and the results
obtained above, we obtain the following theorem.

\begin{theorem} Assume that the antisymmetric matrix $\mathbb{N}$ has corank $d$. Then the Bloch-Iserles system $\dot X(t)=[X(t)^2,\mathbb{N}]$ describes geodesic motion on the generalized Jacobi group $\operatorname{Jac}(2k,d,-\bar{\mathbb{N}}^{-1})$ relative to the left invariant metric given at the identity by the Lagrangian
\[
\ell( \mathcal{S} , \mathcal{A} , \mathcal{B} )=\frac{1}{2}\operatorname{Tr}(\mathcal{S}^T(-\bar{\mathbb{N}}^{-2})\mathcal{S})+\operatorname{Tr}\left(\mathcal{A}^T(-\bar{\mathbb{N}}^{-2})\mathcal{A}\right)+2\operatorname{Tr}(\mathcal{B}^2).
\]
\end{theorem}
This theorem concludes our discussion about how the Bloch-Iserles
system describes a geodesic flow on the generalized Jacobi group.
The next section will focus on the Lie-Poisson bracket on
$\mathfrak{jac}^*(\R^{2k},-\bar{\mathbb{N}}^{-1};\mathbb{R}^d)$ and
will explore its underlying momentum maps.

\subsection{The Hamiltonian structure and its properties}

This section investigates the Lie-Poisson structure of the
Bloch-Iserles system, starting from its underlying Jacobi group.
We first compute the expression of the Lie-Poisson bracket on the Jacobi group and rewrite the Bloch-Iserles system in terms of the variables $(\mathcal{X},\mathcal{K},\mathcal{M})\in\mathfrak{jac}^*(\mathbb{R}^{2k},-\bar{\mathbb{N}}^{-1};\mathbb{R}^d)$. Then we shall show how this Poisson bracket can be simplified (\textit{untangled}). The latter process is made possible by the semidirect-product structure of the Jacobi group, so that reduction by stages \cite{MaMiOrPeRa2007} (see Appendix \ref{RedByStages}) can be applied directly. The simplified Poisson structure is obtained by simple application of a Poisson diffeomorphism which takes the Poisson manifold $\mathfrak{jac}^*(\mathbb{R}^{2k},-\bar{\mathbb{N}}^{-1};\mathbb{R}^d)$ into the direct-product Poisson manifold $\mathfrak{sp}^*(\mathbb{R}^{2k},-\bar{\mathbb{N}}^{-1})\times\mathfrak{h}^*(\mathbb{R}^{2k},-\bar{\mathbb{N}}^{-1})$. Besides recovering an important class of Casimir functions, this process also provides us with a family of momentum map solutions, whose physical explanation will be clarified later in terms of the Klimontovich solution of the underlying Vlasov equation.

\subsubsection{The Lie-Poisson bracket} 
We start our discussion by presenting the Lie-Poisson bracket associated to the generalized Jacobi group, which governs Bloch-Iserles dynamics. 
The (\text{left}) Lie-Poisson bracket on the dual Lie algebra $\mathfrak{jac}(\mathbb{R}^{2k},-\bar{\mathbb{N}}^{-1};\mathbb{R}^d)^*$ is given by
\begin{align}\label{LPGenBI}
\{f,g\}_-(\mathcal{X},\mathcal{K},\mathcal{M})&=-\left\langle(\mathcal{X},\mathcal{K},\mathcal{M}),\left[\left(\frac{\delta f}{\delta\mathcal{X}},\frac{\delta f}{\delta\mathcal{K}},\frac{\delta f}{\delta\mathcal{M}}\right),\left(\frac{\delta g}{\delta\mathcal{X}},\frac{\delta g}{\delta\mathcal{K}},\frac{\delta g}{\delta\mathcal{M}}\right)\right]\right\rangle\nonumber\\
&=-\operatorname{Tr}\left(\mathcal{X}\left[\frac{\delta f}{\delta\mathcal{X}},\frac{\delta g}{\delta\mathcal{X}}\right]\right)
-\operatorname{Tr}\left(\mathcal{K}\left(\frac{\delta f}{\delta\mathcal{X}}\frac{\delta g}{\delta\mathcal{K}}-\frac{\delta g}{\delta\mathcal{X}}\frac{\delta f}{\delta\mathcal{K}}\right)\right)
\\
&\qquad\qquad\qquad-\operatorname{Tr}\left(\mathcal{M}\left(\frac{\delta f}{\delta\mathcal{K}}\right)^{\!T\!\!}(-\bar{\mathbb{N}}^{-1})\,\frac{\delta g}{\delta\mathcal{K}}\right)\nonumber,
\end{align}
where $(\mathcal{X},\mathcal{K},\mathcal{M})\in \mathfrak{sp}(\mathbb{R}^{2k},-\bar{\mathbb{N}}^{-1})\times\mathcal{M}_{d\times 2k}\times\operatorname{Sym}(\mathbb{R}^d)$ and we used the expression \eqref{Lie_bracket_Jac}
for the Lie bracket on the Lie algebra of the generalized Jacobi group. In few cases, it will be convenient to use the notation $\mathfrak{g}^*_\pm$ to distinguish between the left ($\mathfrak{g}^*_-$) and right ($\mathfrak{g}^*_+$) Lie-Poisson brackets accompanying the dual of an arbitrary Lie algebra $\mathfrak{g}$.

Given a Hamiltonian $h=h(\mathcal{X},\mathcal{K},\mathcal{M})$, the associated Lie-Poisson equations are
\begin{equation}\label{LP-equations-on-Jac}
\left\{
\begin{array}{l}
\displaystyle\vspace{0.2cm}\dot{\mathcal{X}}=\left[\mathcal{X},\frac{\delta h}{\delta \mathcal{X}}\right]-\frac{\delta h}{\delta\mathcal{K}}\diamond\mathcal{K}\\
\displaystyle\vspace{0.2cm}\dot{\mathcal{K}}=\mathcal{K}\frac{\delta h}{\delta\mathcal{X}}-\mathcal{M}\left(\frac{\delta h}{\delta\mathcal{K}}\right)^{\!\!T\!}\bar{\Bbb{N}}^{-1}\\
\dot{\mathcal{M}}=0
\end{array}
\right.
\end{equation}
where the diamond operator $\diamond:\mathcal{M}_{2k\times d}\times \mathcal{M}_{d\times 2k}\rightarrow\mathfrak{sp}(\mathbb{R}^{2k},-\bar{\mathbb{N}}^{-1})^*=\mathfrak{sp}(\mathbb{R}^{2k},-\bar{\mathbb{N}}^{-1})$ is defined by the condition
\[
\langle\mathcal{A}\diamond\mathcal{K},\mathcal{S}\rangle=\left\langle
\mathcal{K},\mathcal{S}\mathcal{A}\right\rangle,\quad\text{for all}\quad \mathcal{S}\in \mathfrak{sp}(\mathbb{R}^{2k},-\bar{\mathbb{N}}^{-1}).
\]
We obtain
\begin{align*}
\langle\mathcal{A}\diamond\mathcal{K},\mathcal{S}\rangle &=\left\langle
\mathcal{K},\mathcal{S}\mathcal{A}\right\rangle ={\rm
Tr}(\mathcal{K}\mathcal{S}\mathcal{A})=
{\rm Tr}\left( \mathcal{A}\mathcal{K}\bar{\Bbb{N}}\bar{\Bbb{N}}^{-1}\mathcal{S}\right)
\\
& ={\rm Tr}\left((\mathcal{A}\mathcal{K}\bar{\Bbb{N}})^{\rm
sym}\,\bar{\Bbb{N}}^{-1}\mathcal{S}\right)=\left\langle\left(\mathcal{A}\mathcal{K}\bar{\Bbb{N}}\right)^{\rm sym}\bar{\Bbb{N}}^{-1},\mathcal{S}\right\rangle,
\end{align*}
where we recalled that $\bar{\Bbb{N}}^{-1}\mathcal{S}$ is symmetric.

If $h$ is the Bloch-Iserles Hamiltonian \eqref{BI_Hamiltonian_on_Jac} then evaluating the derivatives $\frac{\delta h}{\delta\mathcal{X}}=\bar{\Bbb{N}}\mathcal{X}\bar{\Bbb{N}}$ and $\frac{\delta h}{\delta\mathcal{K}}=-\frac12\bar{\Bbb{N}}^2\mathcal{K}^T$
yields the explicit equations
\begin{equation}\label{BIEQNS}
\left\{
\begin{array}{l}
\displaystyle\vspace{0.2cm}\dot{\mathcal{X}}=\left[\mathcal{X},\bar{\Bbb{N}}\mathcal{X}\bar{\Bbb{N}}\right]+\frac12\left(\bar{\Bbb{N}}^2\mathcal{K}^T\mathcal{K}\bar{\Bbb{N}}\right)^{\rm sym}\bar{\Bbb{N}}^{-1}\\
\displaystyle\vspace{0.2cm}\dot{\mathcal{K}}=\mathcal{K}\,\bar{\Bbb{N}}\mathcal{X}\bar{\Bbb{N}}+\frac12\,\mathcal{M}\mathcal{K}\bar{\Bbb{N}}\\
\dot{\mathcal{M}}=0
\,.
\end{array}
\right.
\end{equation}
which are the Bloch-Iserles equations on $\mathfrak{jac}(\Bbb{R}^{2k},-\Bbb{N}^{-1},\Bbb{R}^d)^*.$

As we have shown above, the Hamiltonian structure of the Bloch-Iserles system is the ordinary Lie-Poisson structure on the dual of the Lie algebra of the Jacobi group ${\rm Jac}(\R^{2k},-\Bbb{N}^{-1};\R^d)$. However, using the special form of this Lie algebra in terms of its semidirect-product structure
\[
\mathfrak{jac}(\R^{2k},-\Bbb{N}^{-1};\R^d)=\mathfrak{sp}(\R^{2k},-\Bbb{N}^{-1})\,\circledS\,\mathfrak{h}(\R^{2k},-\Bbb{N}^{-1};\R^d)\,,
\]
the next section will show that this Poisson structure can be simplified by using a Poisson diffeomorphism called the untangling map. This map can be naturally introduced in the context of \textit{Poisson reduction by stages}, see \cite{MaMiOrPeRa2007} and Appendix \ref{RedByStages}.
This approach has the advantage of exhibiting the intermediate Poisson manifold $T^*\mathrm{Sp}\times\mathfrak{h}^*$ which is of the form $T^*G\times P$, where $P$ is a Poisson manifold. In this case, there is a procedure to simplify (untangle) the reduced Poisson bracket.

\subsubsection{The untangling momentum map}

The Poisson bracket \eqref{LPGenBI} possesses a special form, whose general setting is best described in \cite{KrMa1987} (see Appendix \ref{untangling-momap-appendix}). As a consequence of this construction, it is possible to transform the Lie-Poisson bracket \eqref{LPGenBI} on $\mathfrak{jac}^*(\R^{2k},\mathbb{J};\R^d)$ into a direct-sum Poisson bracket on $\mathfrak{sp}^*(\R^{2k},\mathbb{J})\oplus\mathfrak{h}^*(\R^{2k},\mathbb{J};\R^d)$. This property involves finding a particular momentum map, which in turn produces interesting considerations on the system, such as Casimir invariants and momentum map solutions. As we shall see, the identification of this momentum map requires restricting $\mathfrak{h}^*(\R^{2k},\mathbb{J};\R^d)$ to an open subset.

From the general theory (see \cite{KrMa1987} and Appendix \ref{untangling-momap-appendix}), the main object is an \emph{untangling momentum map}
$\mathfrak{h}^*(\R^{2k},\mathbb{J};\R^d)\to\mathfrak{sp}^*(\R^{2k},\mathbb{J})$
relative to the induced action of the symplectic group  on
$\mathfrak{h}^*(\R^{2k},\mathbb{J};\R^d)$ , which is given by
\begin{equation}\label{action_Sp_on_h}
(\mathcal{K},\mathcal{M})\mapsto (\mathcal{K}g^{-1},\mathcal{M}).
\end{equation}
From the abstract point of view, this is the representation of the Lie group $G={\rm Sp}$
on the Poisson manifold $\mathfrak{h}^*$. This representation is induced by the homomorphism action of ${\rm Sp}$ on $\rm H$, which
 produces the semidirect-product structure ${\rm Sp}\,\circledS\,{\rm H}$.
The associated infinitesimal generator of
$\mathcal{S}\in\mathfrak{sp}(\mathbb{R}^{2k},\mathbb{J})$ is
\[
\mathcal{S}_{\mathfrak{h}^*}(\mathcal{K},\mathcal{M})=(-\mathcal{K}\mathcal{S},0).
\]
Consider the Lie-Poisson structure on the dual of the generalized Heisenberg algebra $\mathfrak{h}(\R^{2k},\mathbb{J};\R^d)$
\[
\{f,g\}_{-}(\mathcal{K,M})=-\operatorname{Tr}\left(\mathcal{M}\left(\frac{\partial f}{\partial \mathcal{K}}\right)^{\!\!T\!}\mathbb{J}\,\frac{\partial g}{\partial \mathcal{K}}\right).
\]
Since the symplectic group acts by homomorphisms, the action is canonical with respect to the Lie-Poisson structure.
In order to show the existence of a momentum map we need to restrict the above bracket to the open subset $\operatorname{det}\mathcal{M}\neq0$ (cf. \cite{BlBrIsMaRa09}). We shall  denote this restriction by
\[
\widetilde{\mathfrak{h}}^*(\R^{2k},\mathbb{J};\R^d)=\mathcal{M}_{d\times 2k}\times\widetilde{{\rm Sym}}(\R^d)\subset \mathfrak{h}^*(\R^{2k},\mathbb{J};\R^d),
\]
where $\widetilde{{\rm Sym}}(\R^d)$ denotes the set of invertible symmetric $d\times d$ matrices (also known as `inertia tensors'). We shall denote by $\{\,,\}_{\,\widetilde{\mathfrak{h}}^*\!}$ the Lie-Poisson bracket restricted to the open subset $\widetilde{\mathfrak{h}}^*(\R^{2k},\mathbb{J};\R^d)$. Also, we shall use the notation $\widetilde{\mathfrak{jac}}^*(\R^{2k},\mathbb{J};\R^d)=\mathfrak{sp}^*(\R^{2k},\mathbb{J};\R^d)\times\widetilde{\mathfrak{h}}^*(\R^{2k},\mathbb{J};\R^d)$. 
A simple verification shows that
\begin{proposition}[Untangling momentum map]\label{untmomap} The map
\begin{align}\nonumber
{\bf J}_{\widetilde{\mathfrak{h}}^*}:\widetilde{\mathfrak{h}}_-^*(\R^{2k},\mathbb{J};\R^d)&\to\mathfrak{sp}^*(\R^{2k},\mathbb{J})
\\
(\mathcal{K},\mathcal{M})&\mapsto -\frac12\,\mathbb{J}^{-1}\,\mathcal{K}^T\mathcal{M}^{-1}\mathcal{K}
\end{align}
is an equivariant momentum map relative to the action \eqref{action_Sp_on_h} of the symplectic group on the open subset $\widetilde{\mathfrak{h}}^*(\R^{2k},\mathbb{J};\R^d)$ of the generalized Heisenberg algebra.
\end{proposition}
\paragraph{Proof.} Recall that the Poisson bracket  on $\widetilde{\mathfrak{h}}^*(\R^{2k},\mathbb{J};\R^d)$ is
\[
\{f,g\}_{\,\widetilde{\mathfrak{h}}^*}(\mathcal{K,M}):=-\operatorname{Tr}\left(\mathcal{M}\left(\frac{\partial f}{\partial \mathcal{K}}\right)^{\!\!T\!}\mathbb{J}\,\frac{\partial g}{\partial \mathcal{K}}\right).
\]
The momentum map property is verified by observing that
\[
\left\{f,\left\langle{\bf
J}_{\widetilde{\mathfrak{h}}^*},\mathcal{S}\right\rangle\right\}_{\widetilde{\mathfrak{h}}^*}(\mathcal{K},\mathcal{M})=-\operatorname{Tr}\left(\mathcal{K\,S}\,\frac{\partial
f}{\partial
\mathcal{K}}\right)=\operatorname{Tr}\left(\frac{\partial
f}{\partial
\mathcal{K}}(-\mathcal{K\,S})\right)=\mathbf{d}f\,\cdot\,\mathcal{S}_{\,\widetilde{\mathfrak{h}}^*\!}(\mathcal{K},\mathcal{M}),
\]
for all functions $f=f(\mathcal{K},\mathcal{M})$, where the first equality follows from the formula
\[
\frac{\partial}{\partial\mathcal{K}}\left\langle{\bf J}_{\widetilde{\mathfrak{h}}^*},\mathcal{S}\right\rangle=-\mathcal{S}\ \mathbb{J}^{-1}\,\mathcal{K}^T\mathcal{M}^{-1}.
\]
The equivariance is verified as follows
\begin{align*}
\mathbf{J}_{\widetilde{\mathfrak{h}}^*}(\mathcal{K}g^{-1},\mathcal{M})=\frac{1}{2}\mathbb{J}^{-1}g^{-T}\mathcal{K}^T\mathcal{M}^{-1}\mathcal{K}g^{-1}=\frac{1}{2}g\,\mathbb{J}^{-1}\mathcal{K}^T\mathcal{M}^{-1}\mathcal{K}g^{-1}=\operatorname{Ad}^*_{g^{-1}}\left(\mathbf{J}_{\widetilde{\mathfrak{h}}^*}(\mathcal{K},\mathcal{M})\right)
\end{align*}
This completes the proof. $\blacksquare$

\bigskip

Thus, upon writing the symplectic form $\mathbb{J}$ in terms of the Poisson matrix operator as ${\mathbb{J}=-\mathbb{N}^{-1}}$, 
we may untangle the restriction to $\widetilde{\mathfrak{jac}}^*(\R^{2k},-\bar{\Bbb{N}}^{-1};\R^d)$ of the Lie-Poisson bracket \eqref{LPGenBI}, according to the following untangling map (see \eqref{untangling_map} in Appendix \ref{untangling-momap-appendix})
\[
u:(\mathcal{X,K,M})\mapsto\left(\mathcal{X}+\mathbf{J}_{\widetilde{\mathfrak{h}}^*}(\mathcal{K},\mathcal{M}),\mathcal{K},\mathcal{M}\right)=\left(\mathcal{X}+\frac12\,\bar{\Bbb{N}}\,\mathcal{K}^T\mathcal{M}^{-1}\mathcal{K},\mathcal{K},\mathcal{M}\right)=:(\mathcal{Y,K,M})
\,.
\]
Then, the untangled Poisson bracket $\{\,,\}_u$ on $\mathfrak{sp}^*(\R^{2k},\mathbb{J})\oplus\widetilde{\mathfrak{h}}^*(\R^{2k},\mathbb{J};\R^d)$ is the direct sum of the Lie-Poisson structures on $\mathfrak{sp}^*(\R^{2k},-\bar{\Bbb{N}}^{-1})$ and $\widetilde{\mathfrak{h}}^*(\R^{2k},-\bar{\Bbb{N}}^{-1};\R^d)$:
\[
\{f,g\}_u(\mathcal{X},\mathcal{K},\mathcal{M})=-\operatorname{Tr}\left(\mathcal{X}\left[\frac{\delta f}{\delta\mathcal{X}},\frac{\delta g}{\delta\mathcal{X}}\right]\right)-\operatorname{Tr}\left(\mathcal{M}\left(\frac{\delta f}{\delta\mathcal{K}}\right)^{\!T\!\!}(-\bar{\mathbb{N}}^{-1})\,\frac{\delta g}{\delta\mathcal{K}}\right),
\]
and the Hamilton's equations are written as
\begin{equation}\label{untangled_Hamilton_equations}
\left\{
\begin{array}{l}
\displaystyle\vspace{0.2cm}\dot{\mathcal{Y}}
=
\left[\mathcal{Y},\frac{\delta h}{\delta \mathcal{Y}}\right]\\
\displaystyle\vspace{0.2cm}\dot{\mathcal{K}}=-\mathcal{M}\left(\frac{\delta h}{\delta\mathcal{K}}\right)^{\!\!T\!}\bar{\Bbb{N}}^{-1}\\
\dot{\mathcal{M}}=0.
\end{array}
\right.
\end{equation}
See Appendix \ref{UntangledBI} for further details on the above equations.
It is easy to recognize how the untangling map untwines the Lie-Poisson bracket, while it may entangle the Hamiltonian in a complicated way (see Appendix \ref{UntangledBI}). However, this map is also important because it explicitly provides Casimir functions for the entangled bracket. Indeed, from the general theory (see corollary 2.3 in \cite{KrMa1987}) one has the following Casimir functions for the restriction of the Bloch-Iserles bracket \eqref{LPGenBI} to $\widetilde{\mathfrak{jac}}^{\,*\!}(\R^{2k},-\bar{\Bbb{N}}^{-1};\R^d)$:
\[
C_j(\mathcal{X,K,M})=C_j\!\left(\mathcal{X}+\frac12\,\bar{\Bbb{N}}\,\mathcal{K}^T\mathcal{M}^{-1}\mathcal{K}\right)
\]
where
\[
C_j(\mathcal{X})=\frac1{2j}\operatorname{Tr}\!\left((\mathcal{X})^{2j}\right)
\ \qquad\ j=1\dots k
\]
is a Casimir function on $\mathfrak{sp}^*(\R^{2k},-\bar{\Bbb{N}}^{-1})$.
 This result naturally recovers a class of Casimirs already found in \cite{BlBrIsMaRa09} by other methods.

\subsubsection{Momentum map solutions of the Bloch-Iserles system}\label{momap_sol_BI}
This section shows another consequence of the untangling momentum
map in Proposition \ref{untmomap}. Indeed, the equivariance property
of this momentum map determines a special class of solutions of the
Bloch-Iserles system.

We first give the formulas for the adjoint and coadjoint action of the generalized Jacobi group. From the group multiplication law on $\operatorname{Jac}(\mathbb{R}^{2k},\mathbb{J};\mathbb{R}^d)$, we obtain the following formula for the adjoint representation:
\[
\operatorname{Ad}_{(g,\mathcal{A},\mathcal{B})}(\mathcal{S}',\mathcal{A}',\mathcal{B}')=\left(g\mathcal{S}'g^{-1},g\mathcal{A}'-g\mathcal{S}'g^{-1}\mathcal{A},\mathcal{B}'+(\mathcal{A}^T\mathbb{J}g\mathcal{A}')^{sym}-\frac{1}{2}\left(\mathcal{A}^T\mathbb{J}g\mathcal{S}'g^{-1}\mathcal{A}\right)^{sym}\right).
\]
\rem{     
\[
\operatorname{Ad}^*_{(g,\mathcal{A},\mathcal{B})}(\mathcal{Y},\mathcal{K},\mathcal{M})=\left(g^{-1}\left(\mathcal{Y}-\mathcal{A}\mathcal{K}-\frac{1}{2}\mathcal{A}\mathcal{M}\mathcal{A}^T\mathbb{J}\right)g,\left(\mathcal{K}+\mathcal{M}\mathcal{A}^T\mathbb{J}\right)g,\mathcal{M}\right).
\]
}        
For the left coadjoint action we have
\[
\operatorname{Ad}^*_{(g,\mathcal{A},\mathcal{B})^{-1}}(\mathcal{Y},\mathcal{K},\mathcal{M})=\left(g\mathcal{Y}g^{-1}+\mathbb{J}^{-1}\left(\mathbb{J}\mathcal{A}\mathcal{K}g^{-1}\right)^{sym}-\frac{1}{2}\mathcal{A}\mathcal{M}\mathcal{A}^T\mathbb{J},\mathcal{K}g^{-1}-\mathcal{M}\mathcal{A}^T\mathbb{J},\mathcal{M}\right).
\]
Thus, there is a natural left action of the Jacobi group on the
Heisenberg Lie algebra defined by
\begin{equation}\label{Jac_on_Heisenberg}
(g,\mathcal{A},\mathcal{B})\cdot (\mathcal{A}',\mathcal{B}'):=\operatorname{Ad}_{(g,\mathcal{A},\mathcal{B})}(0,\mathcal{A}',\mathcal{B}')=\left(0,g\mathcal{A}',\mathcal{B}'+(\mathcal{A}^T\mathbb{J}g\mathcal{A}')^{sym}\right)
\end{equation}
whose dual left action reads
\begin{equation}\label{Jac_on_dualHeisenberg}
(g,\mathcal{A},\mathcal{B})\cdot(\mathcal{K},\mathcal{M})=\left(\mathcal{K}g^{-1}-\mathcal{M}\mathcal{A}^T\mathbb{J},\mathcal{M}\right).
\end{equation}
These observations produce the following result:
\begin{proposition} The action \eqref{Jac_on_dualHeisenberg} is canonical with respect to the Lie-Poisson bracket on the dual of the generalized Heisenberg Lie algebra. Moreover, when restricted to the open subset $\tilde{\mathfrak{h}}^*(\mathbb{R}^{2k},\mathbb{J};\mathbb{R}^d)$, it admits the equivariant momentum map
\begin{align}\nonumber
\Psi:\widetilde{\mathfrak{h}}_+^*(\R^{2k},\mathbb{J};\R^d)&\to\mathfrak{jac}^*(\R^{2k},\mathbb{J};\R^d)
\\
(\mathcal{K},\mathcal{M}) & \mapsto
\left(\frac12\,\mathbb{J}^{-1}\,\mathcal{K}^T\mathcal{M}^{-1}\mathcal{K},\,\mathcal{K},\,\mathcal{M}\right),
\label{klim}
\end{align}
where the index $+$ denotes the right Lie-Poisson bracket on $\mathfrak{h}^*(\R^{2k},\mathbb{J};\R^d)$.
\end{proposition}
\paragraph{Proof.} Since $\{0\}\times\mathfrak{h}(\mathbb{R}^{2k},\mathbb{J};\mathbb{R}^d)$ is a Lie subalgebra of $\mathfrak{jac}(\mathbb{R}^{2k},\mathbb{J};\mathbb{R}^d)$, the action \eqref{Jac_on_Heisenberg} defines a Lie algebra isomorphism. Therefore, its dual action is canonical with respect to the Lie-Poisson structure on the dual.
We now show the momentum map property. Since the infinitesimal generator reads
\[
(\mathcal{S},\mathcal{A},\mathcal{B})_{\widetilde{\mathfrak{h}}^*}(\mathcal{K},\mathcal{M})=(-\mathcal{K}\mathcal{S}-\mathcal{M}\mathcal{A}^T\mathbb{J},0),
\]
we get
\[
\mathbf{d}f\,\cdot\,(\mathcal{S},\mathcal{A},\mathcal{B})_{\widetilde{\mathfrak{h}}^*}(\mathcal{K},\mathcal{M})=-\operatorname{Tr}\left(\frac{\partial
f}{\partial\mathcal{K}}\mathcal{K}\mathcal{S}\right)-\operatorname{Tr}\left(\frac{\partial
f}{\partial\mathcal{K}}\mathcal{M}\mathcal{A}^T\mathbb{J}\right).
\]
On the other hand, denoting
$\psi=\left\langle\Psi,(\mathcal{S},\mathcal{A},\mathcal{B})\right\rangle$,
we have
\[
\left\{f,\left\langle\Psi,(\mathcal{S},\mathcal{A},\mathcal{B})\right\rangle\right\}=\operatorname{Tr}\left(\mathcal{M}\left(\frac{\partial
f}{\partial \mathcal{K}}\right)^{\!\!T\!}\mathbb{J}\,\frac{\partial
\psi}{\partial
\mathcal{K}}\right)=-\operatorname{Tr}\left(\frac{\partial
f}{\partial \mathcal{K}}\,\mathcal{M}\left(\frac{\partial
\psi}{\partial \mathcal{K}}\right)^{\!\!T\!}\mathbb{J}\right).
\]
where
\[
\frac{\partial \psi}{\partial \mathcal{K}}=\frac{\partial }{\partial
\mathcal{K}}\left\langle\Psi,(\mathcal{S},\mathcal{A},\mathcal{B})\right\rangle=\mathcal{S}\mathbb{J}^{-1}\mathcal{K}^T\mathcal{M}^{-1}+\mathcal{A}.
\]
The map
$\Psi(\mathcal{K},\mathcal{M})=\left(\frac{1}{2}\mathbb{J}^{-1}\mathcal{K}^T\mathcal{M}^{-1}\mathcal{K},\mathcal{K},\mathcal{M}\right)$
is therefore a momentum map.
The equivariance follows from a direct computation. On one hand we have
\begin{align*}
&\Psi((g,\mathcal{A},\mathcal{B})\cdot (\mathcal{K},\mathcal{M}))=\Psi(\mathcal{K}g^{-1}-\mathcal{M}\mathcal{A}^T\mathbb{J},\mathcal{M})\\
&=\left(\frac{1}{2}\mathbb{J}^{-1}g^{-T}\mathcal{K}^T\mathcal{M}^{-1}\mathcal{K}g^{-1}-\frac{1}{2}\mathbb{J}^{-1}g^{-T}\mathcal{K}^T\mathcal{A}^T\mathbb{J}+\frac{1}{2}\mathcal{A}\mathcal{K}g^{-1}-\frac{1}{2}\mathcal{A}\mathcal{M}\mathcal{A}^T\mathbb{J},\mathcal{K}g^{-1}-\mathcal{M}\mathcal{A}^T\mathbb{J},\mathcal{M}\right).
\end{align*}
On the other hand, we have
\begin{align*}
&\operatorname{Ad}^*_{(g,\mathcal{A},\mathcal{B})^{-1}}(\Psi(\mathcal{K},\mathcal{M}))=\operatorname{Ad}^*_{(g,\mathcal{A},\mathcal{B})^{-1}}\left(\frac{1}{2}\mathbb{J}^{-1}\mathcal{K}^T\mathcal{M}^{-1}\mathcal{K},\mathcal{K},\mathcal{M}\right)\\
&=\left(\frac{1}{2}g\mathbb{J}^{-1}\mathcal{K}^T\mathcal{M}^{-1}\mathcal{K}g^{-1}+\mathbb{J}^{-1}\left(\mathbb{J}\mathcal{A}\mathcal{K}g^{-1}\right)^{sym}-\frac{1}{2}\mathcal{A}\mathcal{M}\mathcal{A}^T\mathbb{J},\mathcal{K}g^{-1}-\mathcal{M}\mathcal{A}^T\mathbb{J},\mathcal{M}\right).
\end{align*}
Now we observe that these expressions are equal. $\qquad\blacksquare$
\begin{remark}[Equivariance of $\Psi$]\normalfont
Note that one can add to $\Psi$ an arbitrary function of
$\mathcal{M}$ and still obtain a momentum map. In particular, the
third component is an arbitrary function of $\mathcal{M}$. However,
in order to have equivariance, we need to make the above choice for
$\Psi$.
\end{remark}

\smallskip
\noindent
Being equivariant, the momentum map $\Psi$ is also a Poisson map, as long as both dual Lie algebras $\widetilde{\mathfrak{h}}^*(\R^{2k},\mathbb{J};\R^d)$ and $\mathfrak{jac}^*(\R^{2k},\mathbb{J};\R^d)$ are endowed with the $(+)$ Lie-Poisson structure (or both have the $(-)$ Lie-Poisson structure).
Therefore we have the following result:
\begin{corollary}
Let $h$ be a Hamiltonian on the dual Jacobi Lie algebra and define
the \textit{collective Hamiltonian} $\tilde h:=h\circ\Psi(\mathcal{K}',\mathcal{M}')$ on
$\widetilde{\mathfrak{h}}^*(\mathbb{R}^{2k},-\bar{\Bbb{N}}^{-1};\mathbb{R}^d)$.
Let $\mathcal{K}'$ and $\mathcal{M}'$ be solutions of the collective
Hamilton's equations
\begin{equation}\label{collective_BI_equations}
\left\{
\begin{array}{l}
\displaystyle\vspace{0.2cm}\dot{\mathcal{K}}'=-\mathcal{M}'\left(\frac{\delta \tilde h}{\delta\mathcal{K}'}\right)^{\!T}\bar{\Bbb{N}}^{-1}\\
\dot{\mathcal{M}}'=0.
\end{array}
\right.
\end{equation}
for $\tilde h$ on
$\widetilde{\mathfrak{h}}^*(\mathbb{R}^{2k},-\bar{\Bbb{N}}^{-1};\mathbb{R}^d)$.
Then
\begin{equation}\label{solution-momentum-map}
(\mathcal{X},\mathcal{K},\mathcal{M})=\left(-\frac12\,\bar{\Bbb{N}}^{-1}\,\mathcal{K}'^T\mathcal{M}'^{-1}\mathcal{K}',\,\mathcal{K}',\,\mathcal{M}'\right)
\end{equation}
is a solution of the Lie-Poisson equations \eqref{LP-equations-on-Jac} on the dual Jacobi Lie
algebra.
\end{corollary}
In the case of the Bloch-Iserles system, replacing \eqref{solution-momentum-map} in the expression \eqref{BI_Hamiltonian_on_Jac} we get the collective Hamiltonian
\[
\tilde h(\mathcal{K}',\mathcal{M}')=h\circ\Psi(\mathcal{K}',\mathcal{M}')=\frac{1}{8}\operatorname{Tr}\!\left(\left(\mathcal{K}'^T\mathcal{M}'^{-1}\mathcal{K}'\,\bar{\mathbb{N}}^2\right)^2\right)+\frac{1}{4}\operatorname{Tr}(\mathcal{K}'(-\bar{\mathbb{N}}^2)\mathcal{K}'^T)+\frac{1}{8}\operatorname{Tr}(\mathcal{M}'^2),
\]
and the Lie-Poisson equations \eqref{BIEQNS} collectivize to
\begin{equation}\label{BI_special_solutions}
\left\{
\begin{array}{l}
\displaystyle\vspace{0.2cm}\dot{\mathcal{K}}'=-\,\frac12\left(\mathcal{K}'\,\bar{\Bbb{N}}^2\mathcal{K}'^T\mathcal{M}'^{-1}-\mathcal{M}'\right)\mathcal{K}'\bar{\Bbb{N}}\\
\dot{\mathcal{M}}'=0
\,,
\end{array}
\right.
\end{equation}
according to \eqref{collective_BI_equations}. 

\smallskip

The existence of momentum map solutions is a special feature of Vlasov moment equations. As explained in Remark \ref{SolMomaps}, these solutions are inherited from the Vlasov dynamics through the corresponding Klimontovich solution. In order to understand the deep nature of this phenomenon, it is useful to study the various geometric structures that are shared between moments and their underlying Vlasov system. The next section addresses these questions for the special case of Bloch-Iserles dynamics on the Jacobi group and its generalized version.

\section{Vlasov and the Jacobi group: subgroup inclusions}

Previous sections showed how the BI system is characterized by a well defined geometric construction based on the Jacobi group. As noticed in remark \ref{Averages}, the BI system also possesses a Vlasov formulation, which arises from a quadratic Vlasov Hamiltonian. In turn, the Vlasov equation possesses another interesting geometric footing in terms of Hamiltonian diffeomorphisms and their central extensions (quantomorphisms, in particular). Then, it becomes a natural question to investigate how much of the geometry of the BI system is shared with  Vlasov dynamics. As we shall see, the Jacobi group underlying BI is a subgroup of the quantomorphism group underlying the Vlasov equation. Even in the most general case of BI dynamics (i.e. $\operatorname{corank}\Bbb{N}>1$), we shall see that the generalized Jacobi group is a natural generalization of the quantomorphism group to the case of manifolds possessing a vector-valued symplectic form (see \cite{Vi2010}).
More particularly, the present section shows that \textit{the moment closures determining BI dynamics correspond to particular subgroup inclusions into the group of quantomorphisms of} $V \times \mathbb{R} $. In the case when $\operatorname{corank}\Bbb{N}=1$, these subgroups are given by $\mathbb{R}$, $\operatorname{Sp}(V,\Omega)$, $\operatorname{H}(V,\Omega)$, and $\operatorname{Jac}(V,\Omega)$. Their subgroup inclusions can be established by using an interesting group isomorphism between quantomorphisms and a central extension of the group of symplectic diffeomorphisms previously constructed in \cite{IsLoMi2006}. The latter involves a group 2-cocycle, which we shall call the \textit{ILM cocycle}. Moreover, the subgroup inclusions mentioned above can be extended to the general case of BI dynamics (i.e. $\operatorname{corank}\Bbb{N}>1$), thereby showing that the generalized Jacobi group is itself a natural generalization of the ILM group cocycle.

\subsection{Background on prequantization central extensions}\label{Background_PCE}

Let $(M,\Omega)$ be a connected symplectic manifold and suppose, for simplicity, that $H^1(M, \mathbb{R}  )=\{0\}$. Denote by $\operatorname{Diff}(M,\Omega)=\{\eta\in\operatorname{Diff}(M)\mid \eta^*\Omega=\Omega\}$ the group of symplectic diffeomorphisms of $M$ (more precisely, we restrict to the connected component of its identity). This group is formally a Lie group whose Lie algebra $\mathfrak{X}(M,\Omega)$ consists of Hamiltonian vector fields, that is, $\mathfrak{X}(M,\Omega)=\{X_h\in\mathfrak{X}(M)\mid h\in \mathcal{F} (M)\}$. There is a natural central extension of the Lie algebra $\mathfrak{X}(M,\Omega)$ associated to the exact sequence
\begin{equation}\label{CE_Lie_algebra}
0\rightarrow\mathbb{R}\rightarrow \mathcal{F}(M)\rightarrow \mathfrak{X}(M,\Omega)\rightarrow 0,
\end{equation}
where the Lie algebra brackets on $\mathcal{F}(M)$ and $\mathfrak{X}(M,\Omega)$ are given by the Poisson bracket and minus the Jacobi-Lie bracket of vector fields, respectively.

When $\Omega$ has an integral cohomology class, then there exists a principal circle
bundle $\pi : P \rightarrow M$ over $M$ and a connection one-form $\theta\in\Omega^1(P)$ such that $\mathbf{d}\theta=\pi^*\Omega$. Such a symplectic manifold is said to be \textit{quantizable}. In this case, the Lie algebra extension \eqref{CE_Lie_algebra} can be integrated into a Lie group central extension
\begin{equation}\label{CE_group}
1\rightarrow S^1\rightarrow \operatorname{Aut}(P,\theta)\rightarrow\operatorname{Diff}(M,\Omega)\rightarrow 1,
\end{equation}
of $\operatorname{Diff}(M,\Omega)$, see \cite{Ko1970,So1970,RaSc1981,Vi1997}, called the \textit{prequantization central
extension}. Here $\operatorname{Aut}(P,\theta)$ denotes the group $\{\varphi\in\operatorname{Diff}(P)\mid \varphi^*\theta=\varphi\}$ of \textit{quantomorphisms} (also called \textit{strict contact transformations}) of $(P,\theta)$ (again, in what follows we restrict to the connected component of the identity). Note that $\theta$ is a contact form on $P$ and that a quantomorphism is necessarily an automorphism of the principal bundle.
In order to make clear the link between the central extensions \eqref{CE_group} and \eqref{CE_Lie_algebra}, we recall that the Lie algebra $\mathfrak{aut}(P,\theta)=\{U\in\mathfrak{X}(P)\mid \pounds_U\theta=0\}$ of the quantomorphism group is isomorphic to $\mathcal{F}(M)$, the Lie algebra isomorphism being given by
\begin{equation}\label{crucial_Lie_algebra_isomorphism}
U\in\mathfrak{aut}(P,\theta)\mapsto -\widetilde{\mathbf{i}_U\theta}\in\mathcal{F}(M),
\end{equation}
where $\widetilde{\mathbf{i}_U\theta}\in\mathcal{F}(M)$ is the function on $M$ induced by the $S^1$-invariant function $\mathbf{i}_U\theta$ on $P$. The inverse reads
\begin{equation}\label{crucial_Lie_algebra_isomorphism_inverse}
f\mapsto \operatorname{Hor}X_f-(f\circ\pi)\zeta,
\end{equation}
where $\zeta$ denotes the Reeb vector field, uniquely determined by the conditions $\mathbf{i}_\zeta\theta=1$, $\mathbf{i}_\zeta\mathbf{d}\theta=0$, and $\operatorname{Hor}X_f$ denotes the horizontal lift of the vector field $X_f$ relative to the connection $\theta$. For a trivial bundle $P=M\times\Bbb{R}$, we have $\operatorname{Hor}X=X\partial_m+\alpha(X)\partial_s$ and $\zeta=-\partial_s$.
\begin{remark}[Quantomorphisms vs automorphisms]\normalfont
When the condition $\phi^*\theta=\theta$ is dropped, one is left with the group $\operatorname{Aut}(P)$ of automorphisms of $P$. In the general case, this larger group has no relation with Vlasov dynamics. However, it is interesting to notice that the Mawxell-Vlasov (and consequently Poisson-Vlasov) systems appeared in \cite{CeHoHoMa98} as coadjoint motion on $\operatorname{Aut}(\Bbb{R}^{2k+1})$, at least for the Vlasov part of those systems. This was done upon introducing a redundancy in the construction of the system. In particular, this redundancy consists in adding the equations for particle trajectories, which in turn are already determined by the Vlasov equation alone.
\end{remark}

\subsubsection{The ILM cocycle and its Lie algebra}\label{Sec:ILM}

In the particular case when the symplectic form is exact, we can choose the trivial principal bundle $P=M\times \mathbb{R}  $ and the connection form $\theta=\alpha -\mathbf{d}s$, where $\Omega=\mathbf{d}\alpha$. In this particular case, $\operatorname{Aut}(P,\theta)$ is diffeomorphic to $\operatorname{Diff}(M,\Omega)\times \mathbb{R}  $, and the prequantization central extension can be described by a $\mathbb{R}$-valued group $2$-cocycle on $\operatorname{Diff}(M,\Omega)$, as shown in \cite{IsLoMi2006}. More precisely, given a point $m_0\in M$, the {\bfi ILM cocycle}, defined in \cite{IsLoMi2006}, is
\begin{equation}\label{ILMcocycle}
B_{m_0}(\eta_1,\eta_2):=\int_{m_0}^{\eta_2(m_0)}\left(\eta_1^*\alpha-\alpha\right),\quad \eta_1,\eta_2\in \operatorname{Diff}(M,\Omega),
\end{equation}
where the integral is taken along a smooth curve connecting the point $m_0$ with the point $\eta_2(m_0)$. The cohomology class of $B_{m_0}$ is independent of the choice of the point $m_0$ and the 1-form $\alpha$ such that $\mathbf{d}\alpha=\Omega$, see Theorem 3.1 in \cite{IsLoMi2006}. We denote by
$\operatorname{Diff}(M,\Omega)\times_{B_{m_0}}\mathbb{R}\ni (\eta,a)$,
the associated central extension of the group $\operatorname{Diff}(M,\Omega)$, whose
group multiplication reads
\[
(\eta_1,a_1)(\eta_2,a_2)=(\eta_1\circ\eta_2,a_1+a_2+B_{m_0}(\eta_1,\eta_2)).
\]
Note that here we choose to work with the real line $\mathbb{R}$ instead of the circle $S^1$.
The group isomorphism between the central extension and the group of quantomorphisms is given by
\begin{equation}\label{group_isom}
(\eta,a)\in\operatorname{Diff}(M,\Omega)\times_B\mathbb{R}\mapsto \varphi_{(\eta,a)}\in \operatorname{Aut}(M\times\mathbb{R},\alpha-\mathbf{d}s),
\end{equation}
where $\varphi_{(\eta,a)}$ is defined by
\begin{equation}\label{tilde_eta}
\varphi_{(\eta,a)}(m,s):=(\eta(m),s+a+\tilde{\eta}(m)),\quad\tilde{\eta}(m):=\int_{m_0}^m\left(\eta^*\alpha-\alpha\right),
\end{equation} 
see \cite{IsLoMi2006}.

\begin{remark}[Point transformations]\normalfont
Among all Hamiltonian diffeomorphisms of an arbitrary cotangent bundle $M=T^*Q$,  point transformations are regarded as cotangent lifts of diffeomorphisms of the base configuration manifold $Q$. These transformations possess the fundamental property of preserving the canonical one-form, i.e.  $\alpha={\bf p}\cdot \mathrm{d}{\bf q}$ on $T^*Q=\Bbb{R}^{2k}$. Therefore, the group cocycle \eqref{ILMcocycle} vanishes on the diffeomorphism group $\operatorname{Diff}(Q)$, when the latter is interpreted as given by point transformations on $T^*Q$. The same argument holds for the group of extended point transformations $\operatorname{Diff}(Q)\,\circledS\,\mathcal{F}(Q)$. Notice that the groups $\operatorname{Diff}(Q)$ and $\operatorname{Diff}(Q)\,\circledS\,\mathcal{F}(Q)$ are the Lie groups underlying coadjoint orbits associated to the integrable Camassa-Holm equation and its two-component extension, respectively \cite{HoTr2009}.
\end{remark}

We now compute the Lie algebra $2$-cocycle on $\mathfrak{X}(M,\Omega)$ induced by the ILM cocycle $B_{m_0}$. Recall that this is given by the general formula
\begin{equation}\label{formula_LA_two_cocycle}
C_{m_0}(X,Y)=\left.\frac{d}{dt}\right|_{t=0}\left.\frac{d}{ds}\right|_{s=0}\left(B_{m_0}(\eta_t,\xi_s)-B_{m_0}(\xi_s,\eta_t)\right),
\end{equation}
where $\eta_t$ and $\xi_s$ are curves in $\operatorname{Diff}(M,\Omega)$ such that $\left.\frac{d}{dt}\right|_{t=0}\eta_t=X$
and  $\left.\frac{d}{ds}\right|_{s=0}\xi_s=Y$.
Also, recall  that the Lie bracket on $\mathfrak{X}(M,\Omega)\times_{C_{m_0}} \mathbb{R}$ is given by
\[
[(X,u),(Y,v)]=\left([X,Y],C_{m_0}(X,Y)\right).
\]

\begin{lemma}\label{LA_cocycle} The Lie algebra two cocycle associated to $B_{m_0}$ is given by
\[
C_{m_0}(X_f,X_g)=\{f_\alpha,g\}(m_0)+\{f,g_\alpha\}(m_0),
\]
where $h_\alpha$ is the function on $M$ defined by 
\[
h_\alpha:=h+\alpha(X_h)
\,.
\]
Note that we have $\mathbf{d}h_\alpha=\pounds_{X_h}\alpha$.
\end{lemma}
\noindent\textbf{Proof.} Let $\eta_t$ and $\xi_s$ in $\operatorname{Diff}(M,\Omega)$ be curves tangent to $X_f$ and $X_g$ at $t=0$. We compute
\begin{align*}
\left.\frac{d}{dt}\right|_{t=0}\left.\frac{d}{ds}\right|_{s=0}B(\eta_t,\xi_s)&=\left.\frac{d}{dt}\right|_{t=0}\left.\frac{d}{ds}\right|_{s=0}\int_{m_0}^{\xi_s(m_0)}\left(\eta_t^*\alpha-\alpha\right)=\left\langle\pounds_{X_f}\alpha,X_g\right\rangle(m_0)\\
&=\left\langle\mathbf{d}f_\alpha,X_g\right\rangle(m_0)=\{f_\alpha,g\}(m_0).
\end{align*}
Using formula \eqref{formula_LA_two_cocycle} we obtain the required result.$\qquad\blacksquare$

\medskip

\noindent
The Lie algebra isomorphism associated to \eqref{group_isom} reads
\[
\mathfrak{aut}(M\times\mathbb{R},\alpha-\mathbf{d}s)\rightarrow\mathfrak{X}(M, \Omega )\times_{C_{m_0}}\mathbb{R},\quad X\partial_m+k\partial_s\mapsto (X,k(m_0)).
\]
Besides the Lie algebra isomorphism $\mathfrak{X}(M,\Omega)\times_{C_{m_0}}\mathbb{R}\simeq \mathfrak{aut}(M\times\mathbb{R},\alpha-\mathbf{d}s)$, there is also a well-defined Lie algebra isomorphism with the space $ \mathcal{F} (M)$ of functions on $M$,
\begin{equation}\label{particular_case_crucial_Lie_algebra_isomorphism_inverse}
h\in \mathcal{F}(M)\mapsto X_h\partial_m+h_\alpha\partial_s\in\mathfrak{aut}(M\times\mathbb{R},\alpha-\mathbf{d}s),
\end{equation}
obtained by particularizing \eqref{crucial_Lie_algebra_isomorphism_inverse} to our case.

\subsubsection{The case of symplectic vector spaces}\label{symplectic_vector_space}

In this section, we specialize the above results to the case where $M$ is the vector space $V=\mathbb{R}^{2k}$ and is endowed with a constant symplectic form $\Omega$. We can suppose that we work in Darboux coordinates, that is, we have $\Omega=\mathbf{d}\mathbf{q}\wedge\mathbf{d}\mathbf{p}$, $\bz=(\mathbf{q},\mathbf{p})\in\mathbb{R}^{2k}$.
We consider the central extension of $\operatorname{Diff}(V,\Omega)$ for the particular choices
\[
m_0:=0\quad\text{and}\quad\alpha=\frac{1}{2}(\mathbf{q}\cdot \mathbf{d}\mathbf{p}-\mathbf{p}\cdot \mathbf{d}\mathbf{q}),
\]
that is, we have
\[
B(\eta_1,\eta_2)=\int_0^{\eta_2(0)}\left(\eta_1^*\alpha-\alpha\right).
\]
In this case, the cocycle $C$ simplifies and we have the following

\begin{theorem} The Lie algebra two cocycle for $(V,\Omega)$ and $m_0=0$ is
\[
C(X_f,X_g)=\{f,g\}(0).
\]
\end{theorem}
\noindent\textbf{Proof.} A direct check shows that for all $f\in\mathcal{F}(V)$, we have
\[
\alpha(X_f)(\bz)=-\frac{1}{2}\mathbf{d}f(\bz)\cdot\bz\quad\text{and}\quad\{\alpha(X_f),g\}(0)=-\frac{1}{2}\{f,g\}(0).
\]
Thus, applying Theorem \ref{LA_cocycle}, we have
\begin{align*}
C(X_f,X_g)&=\{f+\alpha(X_f),g\}(0)+\{f,g+\alpha(X_g)\}(0)\\
&=\frac{1}{2}\{f,g\}(0)+\frac{1}{2}\{f,g\}(0)=\{f,g\}(0).\qquad\blacksquare
\end{align*}

\medskip

The Lie algebra bracket on $\mathfrak{X}(V,\Omega)\times_C\mathbb{R}$ is thus given by
\[
[(X_f,u),(X_g,v)]=\left(X_{\{f,g\}},\{f,g\}(0)\right),
\]
and the Lie algebra isomorphisms $\mathcal{F}(V)\rightarrow\mathfrak{aut}(V\times\mathbb{R},\alpha-\mathbf{d}s)\rightarrow \mathfrak{X}(V,\Omega)\times_C\mathbb{R}$
simplifies to $h\mapsto X_h\partial_m+h_\alpha\partial_s\mapsto (X_h,h(0))$.

\subsection{Remarkable subgroups and Vlasov moment closures}\label{remarkable_subgroup}

In this section we show how the groups $\mathbb{R}$, $\operatorname{Sp}(V,\Omega)$, $\operatorname{H}(V,\Omega)$, and $\operatorname{Jac}(V,\Omega)$ are naturally subgroups of the group of quantomorphisms of $V\times\mathbb{R}$, by exhibiting the concrete expression of group inclusions. For each cases, we compute the induced Lie algebra homomorphisms and we show that these subgroups correspond to moment closures of the Vlasov system on $\mathcal{F}(V)^*$. For each examples, we will use the Lie algebra isomorphism
\[
\mathfrak{X}(V,\Omega)\times _C\mathbb{R}\rightarrow \mathcal{F}(V),\quad (X_f,u)\mapsto u+f-f(0).
\]
As we shall see, the groups above are more naturally included in $\operatorname{Diff}(V,\Omega)\times _B\mathbb{R}$ rather than in the group $\operatorname{Aut}(V\times\mathbb{R},\theta)$ of quantomorphisms.

\subsubsection{Translations on $\mathbb{R}$} 
There is a natural inclusion of the real line into $\operatorname{Diff}(V,\Omega)\times_B\mathbb{R}$ given by
\[
\mathbb{R}\hookrightarrow \operatorname{Diff}(V,\Omega)\times_B\mathbb{R},\quad a\mapsto (id,a).
\]
This inclusion is a group homomorphism since $(id,a)(id,b)=(id,a+b+B(id,id))=(id,a+b)$.
The corresponding quantomorphism corresponds to translation by $a$ on the $\mathbb{R}$ factor:
\[
(\bz,s)\mapsto\varphi_{(id,a)}(\bz,s)=(\bz,s+a),
\]
since $\tilde{id}=0$ (see the notation $\tilde{\eta}$ introduced in \eqref{tilde_eta}). 
Taking the tangent map at the identity, we get the Lie algebra inclusion
\[
i_\mathbb{R}:\mathbb{R}\hookrightarrow \mathfrak{X}(V,\Omega)\times _C\mathbb{R}\rightarrow \mathcal{F}(V),\quad u\mapsto (0,u)\mapsto (\bz\mapsto u),
\]
where the last expression means the function with constant value $u$, and the Lie bracket on $\mathbb{R}$ vanishes. The dual map to $i_\mathbb{R}$ is
\[
i^*_\mathbb{R}:\mathcal{F}(V)^*\rightarrow\mathbb{R}^*,\quad i^*_\mathbb{R}(f)=\int_Vf(\bz)\mathbf{d}\bz,
\]
and thus recover the moment $X_0(f)$.

\subsubsection{Linear symplectic transformations} There is a natural inclusion of the symplectic group into $\operatorname{Diff}(V,\Omega)\times_B\mathbb{R}$ given by
\[
\operatorname{Sp}(V,\Omega)\hookrightarrow \operatorname{Diff}(V,\Omega)\times_B\mathbb{R},\quad g\mapsto (g,0).
\]
This inclusion is a group homomorphism since, using $B|_{\operatorname{Sp}(V,\Omega)}=0$, we have $(g_1,0)(g_2,0)=(g_1g_2,B(g_1,g_2))=(g_1g_2,0)$.
The corresponding quantomorphism reads
\[
(\bz,s)\mapsto \varphi_{(g,0)}(\bz,s)=(g\bz,s),
\]
since $\tilde g=0$. Taking the tangent map at the identity, we get the Lie algebra inclusion
\[
i_\mathfrak{sp}:\mathfrak{sp}(V,\Omega)\hookrightarrow \mathfrak{X}(V,\Omega)\times _C\mathbb{R}\rightarrow \mathcal{F}(V),
\]
\[
A\mapsto (\bz\mapsto A\bz,0)\mapsto \left(\bz\mapsto -\frac{1}{2}\operatorname{Tr}\left(\Omega A \bz^{\otimes 2}\right)\right),
\]
where $\Omega$ denotes the matrix of the symplectic form,  $\mathfrak{sp}(V,\Omega)$ is endowed with the usual Lie bracket of matrices, and $\mathcal{F}(V)$ is endowed with the Lie bracket given by the canonical Poisson structure.
The last expression is obtained by noticing that $\bz\mapsto A\bz$ is the Hamiltonian vector field associated to the function
\[
f(\bz)=-\frac{1}{2}\operatorname{Tr}\left(\Omega A \bz^{\otimes 2}\right)=\frac{1}{2}\operatorname{Tr}\left(A^T\Omega\bz^{\otimes 2}\right).
\]
Indeed, we have $\mathbf{d}f(\bz)=-\bz^T\Omega A$ and therefore $X_f(\bz)=-\Omega^{-1}\left(\mathbf{d}f(\bz)\right)^T=A\bz$.
The formula
\[
i_\mathfrak{sp}(A)(\bz)=\frac{1}{2}\operatorname{Tr}\left(A^T\Omega\bz^{\otimes 2}\right),
\]
for the Lie algebra inclusion tells us that the appropriate isomorphism between $\operatorname{Sym}(V)$ and $\mathfrak{sp}(V,\Omega)$ is given by $S=A^T\Omega$, that is,
\[
\Gamma:S\in \operatorname{Sym}(V)\mapsto\Gamma(S):=-\Omega^{-1}S=\mathbb{N}S=A\in \mathfrak{sp}(V,\Omega),
\]
where $\mathbb{N}$ is the Poisson tensor associated to $\Omega$. We thus recover the isomorphism \eqref{Lie_algebra_isom} used above.
Identifying $\mathfrak{sp}(V,\Omega)^*$ with $\mathfrak{sp}(V,\Omega)$ via the Killing form, the dual homomorphism is
\[
i^*_\mathfrak{sp}:\mathcal{F}(V)^*\rightarrow\mathfrak{sp}(V,\Omega)^*,\quad i^*_\mathfrak{sp}(f)=-\frac{1}{2}\int_Vf(\bz)\bz^{\otimes2}\Omega\mathbf{d}\bz.
\]
Note that $i^*_\mathfrak{sp}$ is consistently related to the second moment $X_2(f)$ via the dual map $\Gamma^*:\mathfrak{sp}(V,\Omega)^*\rightarrow \operatorname{Sym}(V)$, $\Gamma^*(\mathcal{X})=-\mathcal{X}\Omega^{-1}=\mathcal{X}\mathbb{N}$, since
\[
\Gamma^*\circ i^*_{ \mathfrak{sp}}(f)=X_2(f).
\]

\subsubsection{Heisenberg group} Consider the natural inclusion
\[
\operatorname{H}(V,\Omega)\hookrightarrow\operatorname{Diff}(V,\Omega)\times_B\mathbb{R},\quad (\mathbf{v},a)\mapsto (\tau_\mathbf{v},a),
\]
where 
\[
\tau_\mathbf{v}(\bz):=\bz+\mathbf{v}
\]
is the translation. We will need the following lemma to show that this inclusion is a group homomorphism (\cite{IsLoMi2006}).

\begin{lemma}\label{lemma_ILM} Consider the diffeomorphisms $\tau_{\mathbf{v}_1}$ and $\tau_{\mathbf{v}_2}$ of $V$. Then
we have
\[
B(\tau_{\mathbf{v}_1},\tau_{\mathbf{v}_2})=\frac{1}{2}\Omega(\mathbf{v}_1,\mathbf{v}_2).
\]
\end{lemma}

Using the lemma, we have
\[
(\tau_{\mathbf{v}_1},a_1)(\tau_{\mathbf{v}_2},a_2)=(\tau_{\mathbf{v}_1}\circ \tau_{\mathbf{v}_1},a_1+a_2+B(\tau_{\mathbf{v}_1},\tau_{\mathbf{v}_1})=\left(\tau_{\mathbf{v}_1+\mathbf{v}_2},a_1+a_2+\frac{1}{2}\Omega(\mathbf{v}_1,\mathbf{v}_1)\right)\\
\]
which proves that the inclusion is a group homomorphism. The corresponding quantomorphism reads
\[
(\bz,s)\mapsto\varphi_{(\tau_\mathbf{v},a)}(\bz,s)=\left(\bz+\mathbf{v},s+a+\frac{1}{2}\Omega(\mathbf{v},\bz)\right),
\]
where we used the identity
\[
\tilde{\tau_\mathbf{v}}(\mathbf{z})=\int_0^\bz\left(\tau_\mathbf{v}^*\alpha-\alpha\right)=B(\tau_\mathbf{v},\tau_\bz)=\frac{1}{2}\Omega(\mathbf{v},\bz).
\]
We thus recover the natural action of the Heisenberg group on $V\times\mathbb{R}$. Taking the tangent map at the identity, we get the Lie algebra inclusion
\[
i_\mathfrak{h}:\mathfrak{h}(V,\Omega)\hookrightarrow\mathfrak{X}(V,\Omega)\times_C\mathbb{R}\rightarrow\mathfrak{aut}(V\times\mathbb{R},\alpha-\mathbf{d}s)\rightarrow\mathcal{F}(V),
\]
\[
(\bw,u)\mapsto \left(\bz\mapsto\bw,u\right)\mapsto\bw\partial_\bz+\left(u+\frac{1}{2}\bw^T\Omega\bz\right)\partial_s \mapsto  \left(\bz\mapsto \bw^T\Omega\bz+u\right),
\]
where we used that $\bz\mapsto \bw$ is the Hamiltonian vector field associated to the function $f(\bz)=\bw^T\Omega\bz$, and that we have the equalities $f_\alpha(\bz)=f(\bz)-\frac{1}{2}\mathbf{d}f(\bz)\cdot \bz=\frac{1}{2}\bw^T\Omega\bz$. Recall that the Lie bracket on $\mathfrak{h}(V,\Omega)$ is
\[
[(\mathbf{v},u),(\mathbf{w},v)]=(0,\Omega(\mathbf{v},\bw)),
\]
and that on $\mathcal{F}(V)$ it is given by the symplectic Poisson bracket. The dual homomorphism is
\[
i^*_\mathfrak{h}:\mathcal{F}(V)^*\rightarrow\mathfrak{h}(V,\Omega)^*,\quad i^*_\mathfrak{h}(f)=\left(-\int_Vf(\bz)\bz^T\Omega\mathbf{d}\bz,\int_Vf(\bz)\mathbf{d}\bz\right)
\]
The formula
\[
i_\mathfrak{h}(\bw,u)(\bz)=\bw^T\Omega\bz+u
\]
for the Lie algebra inclusion tells us that the appropriate identification between $\mathfrak{h}(V,\Omega)$ and the Lie algebra encoding the moments $X_0, X_1$ is
\[
(S_1,S_0)\in V^*\times \mathbb{R}\rightarrow (\bw,u):=\gamma(S_1,S_0)=\left(-\Omega^{-1}S_1^T,S_0\right)\in \mathfrak{h}(V,\Omega)
\]
with dual map
\[
\gamma^*:(B,b)\in \mathfrak{h}(V,\mathbb{R})^*\rightarrow (X_1,X_0)=\beta^*(B,b)=\left(\Omega^{-1}B^T,b\right)\in V\times \mathbb{R}.
\]
We consistently have the relation
\[
\gamma^*\circ i^*_\mathfrak{h}(f)=(X_1(f),X_0(f)).
\]

\subsubsection{Jacobi group} Consider the inclusion
\[
\operatorname{Jac}(V,\Omega)=\operatorname{Sp}(V,\Omega)\,\circledS\,\operatorname{H}(V,\Omega)\hookrightarrow \operatorname{Diff}(V,\Omega)\times_B\mathbb{R},\quad (g,\mathbf{v},a)\mapsto (\tau_\mathbf{v}\circ g,a).
\]
We will need the following lemma to show that this inclusion is a group homomorphism.

\begin{lemma} Given $g_1, g_2\in\operatorname{Sp}(V,\Omega)$ and $\mathbf{v}_1, \mathbf{v}_2\in V$, we have
\[
B\left(\tau_{\mathbf{v}_1}\circ g_1, \tau_{\mathbf{v}_2}\circ g_2\right)=\frac{1}{2}\Omega(\mathbf{v}_1,g_1\mathbf{v}_2).
\]
\end{lemma}
\textbf{Proof.} Using the group 2-cocycle property of $B$, we get
\begin{align*}
B\left(\tau_{\mathbf{v}_1}\circ g_1, \tau_{\mathbf{v}_2}\circ g_2\right)&=B\left(\tau_{\mathbf{v}_1}\circ g_1, \tau_{\mathbf{v}_2}\right)+B\left(\tau_{\mathbf{v}_1}\circ g_1\circ\tau_{\mathbf{v}_2}, g_2\right)-B\left(\tau_{\mathbf{v}_2},g_2\right)\\
&=B\left(\tau_{\mathbf{v}_1}\circ g_1, \tau_{\mathbf{v}_2}\right)=\int_0^{\mathbf{v}_2}\left(g_1^*\tau_{\mathbf{v}_1}^*\alpha-\alpha\right)=\int_0^{g_1\mathbf{v}_2}\tau_{\mathbf{v}_1}^*\alpha-\int_0^{\mathbf{v}_2}\alpha\\
&=B\left(\tau_{\mathbf{v}_1},\tau_{g_1\mathbf{v}_2}\right)-0=\frac{1}{2}\Omega(\mathbf{v}_1,g_1\mathbf{v}_2).\qquad\blacksquare
\end{align*}

Recall that the group multiplication on $\operatorname{Jac}(V,\Omega)$ reads
\[
(g_1,\mathbf{v}_1,a_1)(g_2,\mathbf{v}_2,a_2)=\left(g_1g_2,\mathbf{v}_1+g_1\mathbf{v}_2,a_1+a_2+\frac{1}{2}\Omega(\mathbf{v}_1,g_1\mathbf{v}_2)\right).
\]
We thus have
\begin{align*}
(\tau_{\mathbf{v}_1}\circ g_1,a_1)(\tau_{\mathbf{v}_2}\circ g_2,a_2)&=\left(\tau_{\mathbf{v}_1}\circ g_1\circ\tau_{\mathbf{v}_2}\circ g_2,a_1+a_2+B\left(\tau_{\mathbf{v}_1}\circ g_1,\tau_{\mathbf{v}_2}\circ g_2\right)\right)\\
&=\left(\tau_{g_1\mathbf{v}_2+\mathbf{v}_1}\circ g_1g_2,a_1+a_2+\frac{1}{2}\Omega\left(\mathbf{v}_1,g_1\mathbf{v}_2\right)\right),
\end{align*}
which corresponds to the group multiplication on the Jacobi group. The injection is therefore a group homomorphism.

The corresponding strict contact transformation is
\[
(\bz,s)\mapsto \varphi_{(\tau_\mathbf{v}\circ g,a)}(\bz,s)=\left(g\bz+\mathbf{v}, s+a+\frac{1}{2}\Omega(\mathbf{v},g\bz)\right),
\]
since
\[
\widetilde{\tau_\mathbf{v}\circ g}(\bz)=\int_0^\bz\left(g^*\tau_\mathbf{v}^*\alpha-\alpha\right)=\int_0^{g\bz}\tau_\mathbf{v}^*\alpha=B(\tau_\mathbf{v},g\bz)=\frac{1}{2}\Omega(\mathbf{v},g\bz).
\]
Taking the tangent map at the identity, we get the Lie algebra inclusion
\begin{equation}\label{i_jac}
i_\mathfrak{jac}:\mathfrak{jac}(V,\Omega)\hookrightarrow\mathfrak{X}(V,\Omega)\times_C\mathbb{R}\rightarrow\mathfrak{aut}(V\times\mathbb{R},\alpha-\mathbf{d}s)\rightarrow\mathcal{F}(V),
\end{equation}
\begin{align*}
(A,\bw,u)\mapsto \left(\bz\mapsto A\bz+\bw,u\right)&\mapsto\left(A\bz+\bw\right)\partial_\bz+\left(u+\frac{1}{2}\bw^T\Omega\bz\right)\partial_s\\
& \mapsto  \left(\bz\mapsto\frac{1}{2}\operatorname{Tr}\left(A^T\Omega \bz^{\otimes 2}\right)+\bw^T\Omega\bz+u\right),
\end{align*}
where we recall that the Lie bracket on $\mathfrak{jac}(V,\Omega)$ is
\begin{align*}
[(A,\mathbf{v},u),(B,\mathbf{w},v)]&=([A,B],A\cdot (\bw,v)-B\cdot (\mathbf{v},u)+[(\mathbf{v},u),(\bw,v)])\\
&=\left([A,B],A\bw-B\mathbf{v},\Omega(\mathbf{v},\mathbf{w})\right).
\end{align*}
As before, the formula for the Lie algebra inclusion tells us that we need to consider the isomorphism
\[
\Gamma:\operatorname{Sym}(V)\times V^*\times\mathbb{R}\rightarrow \mathfrak{jac}(V,\Omega),
\]
given by
\[
(S_2,S_1,S_0)\rightarrow (A,\bw,u):=(-\Omega^{-1}S_2,-\Omega^{-1}S_1^T,S_0).
\]
We thus recover the isomorphism \eqref{Gamma_corankone}. As we have seen, $\Gamma$ endows $\operatorname{Sym}(V)\times V^*\times\mathbb{R}$ with the Lie bracket
\[
\left[(S_2,S_1,S_0),(\bar S_2,\bar S_1,\bar S_0)\right]=\left([S_2,\bar S_2]_{\mathbb{N}},S_1\mathbb{N}\bar S_2-\bar S_1\mathbb{N}S_2, S_1\mathbb{N}\bar S_1^T\right),
\]
where $\mathbb{N}=-\Omega^{-1}$ is the Poisson tensor associated to the symplectic form $\Omega$.
The dual map $\Gamma^*:\mathfrak{jac}(V,\Omega)^*\rightarrow\operatorname{Sym}(V)^*\times V\times \mathbb{R}$ consistently relates the dual homomorphism $i^*_\mathfrak{jac}:\mathcal{F}(V)^*\rightarrow\mathfrak{jac}(V,\Omega)^*$
\[
i^*_\mathfrak{jac}(f)(\bz)=\left(-\frac{1}{2}\int_Vf(\bz)\bz^{\otimes2}\Omega\mathbf{d}\bz,-\int_Vf(\bz)\bz^T\Omega\mathbf{d}\bz,\int_Vf(\bz)\mathbf{d}\bz\right)
\]
to the three first moments $(X_2(f), X_1(f),X_0(f))$,
via the relation
\[
\Gamma^*\circ i_{\mathfrak{jac}}^*(f)=\left( X_2(f), X_1(f),X_0(f) \right) .
\]

The main results of this section are summarized in the following theorem.

\begin{theorem} The first three moments $X _0 (f), X _1 (f), X _2 (f)$ of the Vlasov dynamics are associated to the subgroup inclusion of the symmetry group of the Bloch-Iserles system into the symmetry group of the Vlasov dynamics, i.e. the inclusion of the Jacobi group into the quantomorphism group. These moments are given by the Poisson map obtained by taking the dual map to the associated Lie algebra inclusion.
\end{theorem}

\subsection{A commuting diagram underlying the Klimontovich map}

As we have just proven, there is a strict relation between the Vlasov kinetic
equation and the Bloch-Iserles system when
$\operatorname{corank}(\Bbb{N})=1$. We now show that, when $\operatorname{corank}(\Bbb{N})=1$, the
momentum map solution of the Bloch-Iserles system obtained in \S\ref{momap_sol_BI} is
naturally associated to the Klimontovich momentum map of the
corresponding Vlasov equation.
The latter is the solution provided
by the single-particle trajectory and it is given explicitly by (cf.
\cite{Kl1967})
\[
f(\bz,t)=\delta(\bz-\bzeta(t))=:\mathbf{J}_{Klim}(\bzeta(t)).
\]
The map $\mathbf{J}_{Klim}:V\rightarrow\mathcal{F}(V)^*$ is the momentum mapping associated to the natural left action of
$(\eta, a)\in \operatorname{Diff}(V,\Omega )\times_B\mathbb{R}$ on $ \boldsymbol{\zeta}\in V$ given by
$\boldsymbol{\zeta}\mapsto \eta(\boldsymbol{\zeta})$. Being equivariant, $ \mathbf{J}_{Klim}$ is a Poisson map relative to the symplectic Poisson structure on $V$ and the $(+)$ Lie-Poisson structure on $ \mathcal{F} (V)^*$.
A review of the momentum map
properties of the Klimontovich solutions was presented in
\cite{HoTr2009}. Note that by applying the moment map $f\mapsto (X _2 (f), X _1  (f), X _0 (f))$  to the Klimontovich solution we get the solution $\left(\frac12\,\bzeta\,\bzeta^T,\bzeta,1\right)$ of the Bloch-Iserles system (in the case when $X_0(f)=1$). This solution corresponds to a momentum map $\mathbf{K}_2:V\to\mathrm{Sym}(V)^*\times V\times\Bbb{R}^*$, whose last two components yield another momentum map $\mathbf{K}_1:V\to V\times\Bbb{R}^*$.
On the other hand, recall from \S\ref{momap_sol_BI} that there is another momentum map solution $\Psi: \tilde{ \mathfrak{h}  }^*( \mathbb{R}  ^{2k}, \Bbb{J} ; \mathbb{R}  ^d ) \rightarrow \mathfrak{jac}^*( \mathbb{R}  ^{2k}, \Bbb{J} ; \mathbb{R}  ^d )$ of the Bloch-Iserles system, associated to the action \eqref{Jac_on_dualHeisenberg} of the Jacobi group. We now show that, when $d=1$, these momentum map solutions are naturally related by an equivariant momentum map $\mathbf{J}_{H}: V \rightarrow \mathfrak{h} ^*( V, \Omega )$ associated to the natural action of the Heisenberg group on $V$.
More precisely, upon recalling Remark \ref{SpecialGamma}, we will show that we have the following commutative diagram
\begin{diagram}
\mathrm{Sym}(V)^*\times V\times\Bbb{R}^*& &  & \lTo{\textstyle\mu} &  &&  V\times\Bbb{R}^*
\\
 & \luTo(3,4)^{\mathbf{K}_2}\luTo(3,2)^{(X_2,X_1,X_0)}& &   &    &\ruTo(3,2)^{(X_1,X_0)} \ruTo(3,4)^{\mathbf{K}_1} &
\\
&&&\mathcal{F}^*(V)&&&
\\
\uTo^{\textstyle\Gamma^*}&&\ldTo(3,4){\textstyle{i}^*_\textit{jac}}&\uTo_{\!\mathbf{J}_\textit{\!Klim}}&\rdTo(3,4){\textstyle{i}^*_\textit{h}}&&\uTo_{\textstyle\Gamma_{\mathfrak{h}}^*}
\\
&& & (V,\Omega) &&&
\\
&&\ldTo(3,2)_{\textstyle\mathbf{J}_\textit{\!Jac}}&& \rdTo(3,2)_{\textstyle\mathbf{J}_\textit{\!H}}&&
\\
\mathfrak{jac}^*(V,\Omega)& & & \lTo^{\textstyle\Psi} & & & \  \widetilde{\mathfrak{h}}^*(V,\Omega)
\end{diagram}
where
\[
\mu(X_1,X_0)=\left(\frac1{2 X_0}X_1^{\otimes 2},X_1,X_0\right)\,.
\]
The upper left triangle commutes by the results obtained in the previous sections. As we just said $ \mathbf{J} _{Klim}$ is the momentum map associated to the left action of
$\operatorname{Diff}(V,\Omega )\times_B\mathbb{R}$ on $V$. When the
subgroup $\operatorname{Jac}(V,\Omega)\subset
\operatorname{Diff}(V,\Omega)\times_B\mathbb{R}$ acts on $V$,
by the subgroup action, i.e. $(g, \mathbf{v} , a) \cdot \boldsymbol{\zeta }= g \boldsymbol{ \zeta }+ \mathbf{v} $, it
produces the momentum map
\[
\mathbf{J}_{Jac}:V\rightarrow\mathfrak{jac}^*(V,\Omega),\quad
\mathbf{J}_{Jac}(\boldsymbol{\zeta})=\left(-\frac{1}{2}\boldsymbol{\zeta}\boldsymbol{\zeta}^T\Omega,-\boldsymbol{\zeta}^T\Omega,1\right).
\]
Therefore, we have the relation
$\mathbf{J}_{Jac}=i^*_{jac}\circ\mathbf{J}_{Klim}$, where
$i_{jac}^*$ is the dual to the Lie algebra inclusion. Note that
$i^*_{jac}$ is itself a momentum mapping relative to the right
coadjoint action of the Jacobi group on $\mathcal{F}(V)^*$. Similarly, when the subgroup $ \operatorname{H}(V, \Omega )$ acts on $V$ by translations $\boldsymbol{\zeta}\mapsto\boldsymbol{\zeta}+{\bf v}$, it produces the momentum map
\[
\mathbf{J}_{H}:V\rightarrow\mathfrak{h}^*(V,\Omega),\quad
\mathbf{J}_{H}(\boldsymbol{\zeta})=\left(-\boldsymbol{\zeta}^T\Omega,1\right)
\]
and we have the relation $\mathbf{J} _H= i_{h}^* \circ \mathbf{J} _{Klim}$, where $i_h$ is the Lie algebra inclusion. Note that both $ \mathbf{J}_H$ and $\mathbf{J} _{Jac}$ are equivariant and hence Poisson with respect to the symplectic Poisson structure on $V$ and the $(+)$ Lie-Poisson structures.
 Besides the trivial relation $\mathbf{J} _H= i^* \circ \mathbf{J} _{Jac}$ between $ \mathbf{J} _H $ and $ \mathbf{J} _{Jac}$ (here $i$ denotes the Lie algebra inclusion $i:\mathfrak{h}(V,\Omega)\to\mathfrak{jac}(V,\Omega)$), we now show that $\mathbf{J} _{Jac}= \Psi\circ \mathbf{J} _H$. Indeed, we have
\begin{align*}
\Psi\circ\mathbf{J} _H(\boldsymbol{\zeta})=\Psi\left(-\boldsymbol{\zeta}^T\mathbb{J},1\right)=&\left(\frac{1}{2}\mathbb{J}^{-1}(-\boldsymbol{\zeta}^T\mathbb{J})^T(-\boldsymbol{\zeta}^T\mathbb{J}),-\boldsymbol{\zeta}^T\mathbb{J},1\right)
\\
=&\left(-\frac{1}{2}\boldsymbol{\zeta}\boldsymbol{\zeta}^T\mathbb{J},-\boldsymbol{\zeta}^T\mathbb{J},1\right)=\mathbf{J}_{Jac}(\boldsymbol{\zeta})
=i^*_{jac}\circ\mathbf{J}_{Klim}(\boldsymbol{\zeta})
.
\end{align*}
This formula relates, in the case $\operatorname{corank}\Bbb{N}=1$, the particular
solutions given by the momentum map $\Psi$ and those obtained by the
Klimontovich solution of the Vlasov equation via the first three
moments.

\subsection{Remarks on the higher corank case}

While the previous sections addressed the question of how the Jacobi group is related to the group underlying Vlasov dynamics, it becomes natural to ask whether similar relations may appear for the generalized Jacobi group, which underlies BI dynamics for $\operatorname{corank}\Bbb{N}>1$. The first problem that we encounter in addressing this question is that there is no Vlasov moment operation producing dynamical quantities in $\mathfrak{jac}^*(\Bbb{R}^{2k},\Bbb{J};\Bbb{R}^d)$. Indeed, such moment quantities should arise from an averaging process involving a momentum map ${\bf J}: (\Bbb{R}^{2k},\Bbb{J})\to\mathfrak{jac}^*(\Bbb{R}^{2k},\Bbb{J};\Bbb{R}^d)$, along the lines described in Remark \ref{Averages}. However, the generalized Jacobi group does not possess a natural action on $ (\Bbb{R}^{2k},\Bbb{J})$, so that the momentum map $\bf J$ is unlikely to exist. On the other hand, the generalized Jacobi group acts naturally on $\mathcal{M}_{2k\times d}$, which in turn is not a symplectic vector space. Thus, we conclude that the case $\operatorname{corank}\Bbb{N}>1$ of BI dynamics does not fit in the phase-space averaging principles outlined in the Remark \ref{Averages}. Consequently, one is led to seek a generalization of Vlasov dynamics which could include the BI system in the most general case. 
This section addresses the question of how one could develop a similar approach as above for the Bloch-Iserles system with an arbitrary $\operatorname{corank}\Bbb{N}=d>1$. In this case, the resulting group of quantomorphisms (generalizing Vlasov dynamics) turns out to be the generalized Jacobi group itself. So, contrarily to the corank one case, there is no generalization of  Vlasov dynamics that comprises the Bloch-Iserles system when $d\geq 2$. Or, in other words, this Vlasov dynamics coincides with the Bloch-Iserles system itself. 

Recall that passing from $d=1$ to an arbitrary corank $d$ amounts in replacing the Jacobi group $ \operatorname{Jac}(V, \Omega )= \operatorname{Sp}(V, \Omega ) \,\circledS\, \operatorname{H}(V, \Omega )$ by the generalized Jacobi group $ \operatorname{Jac}(V, \Omega ;W)= \operatorname{Sp}(V, \Omega ) \,\circledS\, \operatorname{H}(V, \Omega;W )$ defined in \S\ref{def_gen_Jac}. Contrarily to $ \operatorname{Jac}(V, \Omega )$, the generalized Jacobi group is not a subgroup of $ \operatorname{Diff}(V, \Omega ) \times _B \mathbb{R}  \simeq \operatorname{Aut}(V \times \mathbb{R}  , \alpha - \mathbf{d} s )$. We thus seek a central extension of the $\operatorname{Diff}(V, \Omega )$ generalizing the ILM central extension such that it has an automorphism interpretation and contains in a natural way the generalized Jacobi group.
We already know that the cocycle involved should coincide with the cocycle of the generalized Heisenberg group, when restricted on $\operatorname{Jac}(V, \Omega ;W)$. Therefore, the base space involved will be $L(W,V)$, the extension is made by the vector space $\operatorname{Sym}(W)$, and the 2-form involved is no more a symplectic form: rather it is a closed vector-valued 2-form. Such an approach should thus be consistent with a generalized version of prequantization central extension by vector spaces. Such a theory indeed exists and has been recently developed in \cite{NeVi2003,Vi2010}, as we now review very briefly.

\subsubsection{Abelian prequantization extension} 
Let $E$ be a vector space and $\omega\in\Omega^2(M,E)$ be a closed $E$-valued two-form on $M$. Suppose for simplicity that  $H^1(M,E)=\{0\}$, or, equivalently $H^1(M)=\{0\}$ and let $\operatorname{Diff}(M,\omega)$ be the connected component of the identity of the group of diffeomorphisms that preserves $\omega$. When the period group $ Z _\omega $ of $ \omega $ is discrete, then there exists an Abelian principal bundle $\pi:P\rightarrow M$ with structure group $A=E/Z _\omega $, and a connection one-form $\theta\in\Omega^1(P,E)$ such that $\mathbf{d}\theta=\pi^*\omega$. In this case, we have a central extension of $\operatorname{Diff}(M,\omega)$ given by
\[
1\rightarrow A\rightarrow \operatorname{Aut}(P,\theta)\rightarrow \operatorname{Diff}(M,\Omega)\rightarrow 1,
\]
as shown in \cite{NeVi2003}, where $ \operatorname{Aut}(P, \theta )$ is the connected component of the identity of the group of automorphisms of the principal bundle that preserve the connection form $ \theta $. The associated exact sequence of Lie algebra reads
\[
0\to E\to \mathfrak{aut}(P,\theta)\to\mathfrak{X}(M,\omega)\to 0.
\]
A very important point for the present discussion lies in the fact that the Lie algebra isomorphism \eqref{crucial_Lie_algebra_isomorphism} is now replaced by the surjective linear map
\begin{equation}\label{crucial_surjective_map}
U\in \mathfrak{aut}(P,\theta)\mapsto -\widetilde{\mathbf{i}_U\theta}\in\mathcal{F}_{adm}(M,E),
\end{equation}
where $\mathcal{F}_{adm}(M,E)$ denotes the space of \textit{admissible functions} \cite{Vi2010} defined by 
\[
\mathcal{F}_{adm}(M,E):=\{ f\in\mathcal{F}(M,E)\mid \mathbf{d} f \in \operatorname{Im}( \omega ^\flat)\}
\,.
\]

Note that for all $f, g\in \mathcal{F} _{adm}(M,E)$, it is possible to define the bracket
\begin{equation}\label{Lie_bracket_d}
\{\!\{f,g\}\!\}:= \omega (X, Y)= \mathbf{d} f(Y)=- \mathbf{d} g (X),
\end{equation} 
where $X$ and $Y$ are arbitrary Hamiltonian vector fields associated to $f$ and $g$, respectively. This bracket is well-defined, since $X$ and $Y$ are determined up to a section of $ \operatorname{ker}( \omega ^\flat)$. The second and third equalities are also well defined since $ \mathbf{d} f|_{\operatorname{ker}( \omega ^\flat )}= \mathbf{d} g|_{\operatorname{ker}( \omega ^\flat) }=0$, since $f$ and $g$ are admissible functions. It can be shown that when $f, g\in \mathcal{F} _{adm}(M,E)$, then $\{\!\{f,g\}\!\}\in \mathcal{F} _{adm}(M,E)$ and $\{\!\{\,,\}\!\}$ is a Lie bracket on $\mathcal{F} _{adm}(M,E)$.

Let us observe that if the closed form $ \omega $ is exact, then the Abelian prequantization central extension can be defined by a global $2$-cocycle, given by the same formula as the ILM cocycle \eqref{ILMcocycle}, although now taking values in $E$. We thus obtain a central extension $\operatorname{Diff}(M, \omega ) \times_{B_{m _0 }} E$, with Lie algebra $ \mathfrak{X}_{ex}(M, \omega )\times_{C_{m _0 }}E= \mathfrak{X}(M, \omega )\times_{C_{m _0 }}E$. The same formula \eqref{group_isom} as above realizes a group isomorphism between $\operatorname{Diff}(M, \omega ) \times_{B_{m _0 }} E$ and the group $ \operatorname{Aut}(M\times E, \alpha - \mathbf{d} e )$ of quantomorphisms of the trivial bundle $M\times E$, where $ \mathbf{d} e\in \Omega ^1 (E,E)$ is
defined by $ \mathbf{d} e(u,v)=v$.

\subsubsection{The space of admissible functions is finite dimensional} Let us now consider the particular case when $M= L(W,V)$, $E= \operatorname{Sym}(W)$ 
and the closed $ \operatorname{Sym}(W)$-valued $2$-form on $L(W,V)$ is defined by
\[
\Omega ^W(A,B):= \left( A^T \Omega B \right) ^{sym}.
\]
This is the natural generalization of the setting developed in \S\ref{symplectic_vector_space} for the corank one case. We will however show that when $d\geq 2$, we get nothing else than the generalized Jacobi group itself.

For an arbitrary $(n \times d)$ matrix $A\in L(W,V)$, we denote by $A _i , i=1,..., d$ its column. We thus have
\[
\Omega ^W(A,B)_{ij} = \frac{1}{2} \left( A _i ^T \Omega B _j + A _j ^T \Omega B _i  \right)
\]
We denote by $ f_{ij} = f_{ji}: L(W,V) \rightarrow \operatorname{Sym}(W)$ the components of a function $f: L(W,V) \rightarrow \operatorname{Sym}(W)$, and by $\frac{\partial f _{ij} }{\partial A_k}$ the partial derivative of $f$ relative to the column $A_k$, so that $ \mathbf{d} f _{ij} (A) \cdot B=\sum_{k=1}^d\frac{\partial f _{ij} }{\partial A_k} \cdot  B _k $. Using these notations we have the following result. (Notice that, in what follows, the summation convention on repeated indices is not used, unless otherwise specified).

\begin{lemma}\label{F_adm}
The space of admissible functions associated to $M=L(W,V)$ and the closed $\operatorname{Sym}(W)$-valued 2-form $\Omega^W$ is
\[
\mathcal{F}_{adm}(L(W,V), \operatorname{Sym}(W))=\left \{ f \left | \;\frac{\partial f _{ij}}{\partial  A_k} = 0,\;\;\text{for all $k\neq i,j$},\quad \frac{\partial f _{ij}}{\partial  A_i}=\frac{1}{2} \frac{\partial f _{jj}}{\partial  A_j},\;\;\text{for all $i\neq j$}\right.\right\}
\]
In particular, the kernel of $ \Omega ^W $ is trivial.\\
-- If $d=\operatorname{dim}(W)=1$, there is no condition and $\mathcal{F}_{adm}(V)= \mathcal{F} (V)$.\\
-- If $d=\operatorname{dim}(W)\geq 2$, then the space of admissible function is finite dimensional, with dimension $\frac{m(m+1)}{2}$, $m= \operatorname{dim}(V)+ \operatorname{dim}(W)$ and consists of functions with matrix elements
\[
f _{ij} (A _i , A _j )=C _{ij} + \varphi _i (A_i)+ \varphi  _j (A _j )+B( A _i , A _j ),
\]
where $\varphi_i$, $i=1,...,d$ are linear forms and $B$ is a symmetric bilinear form.
\end{lemma}
\textbf{Proof.} Recall that a function $f$ is admissible if and only if there exists a vector field $X\in \mathfrak{X}(L(W,V))$ such that $ \Omega ^W (X(A),B)= \mathbf{d} f(A) \cdot B$, for all $A, B\in L(W,V)$. Writing this condition for the matrix elements $(i,j)$, we get
\[
\frac{1}{2} (X_i^T\Omega B_j+X_j^T\Omega B_i)=\sum_{k=1}^d \frac{\partial f _{ij} }{\partial A _k }\cdot B _k.
\]
So, we immediately get the first condition $\frac{\partial f _{ij}}{\partial  A_k} = 0$ for all $k\neq i,j$, for all $i,j$. We also have $ \frac{\partial f_{ii}}{\partial A_i}=X_i^T \Omega $, for all $i$ and $ \frac{\partial f_{ij}}{\partial A_i}= \frac{1}{2} X _i^T  \Omega $, for all $i\neq j$. This proves the second condition.

It is clear that when $d=1$ both conditions are empty and so all functions are admissible in this case, consistently with the statements in \S\ref{Background_PCE}.

We now suppose $ d\geq 2$. From the first condition, $f_{ij}$ depends on the variables $A _i $ and $A _j $ only. Thus for $i\neq j$, we get
\[
\frac{\partial ^2f_{ij}}{\partial A_i^2}=\frac{1}{2} \frac{\partial ^2 f_{jj}}{\partial A_i \partial A_j}  =0\quad\text{and}\quad \frac{\partial ^2f_{ij}}{\partial A_j^2}=\frac{1}{2} \frac{\partial ^2 f_{ii}}{\partial A_j \partial A_i}  =0.
\]
This means that $f_{ij}$ is necessarily of the form $f_{ij}(A_i, A_j)= C_{ij}+ \varphi_{ij} (A_i)+\psi_{ij}(A_j)+ B_{ij}(A_i,A_j)$,
where $ \varphi _{ij} $ and $\psi_{ij} $ are linear forms and $B_{ij} $ is a bilinear form. Using the fact that the second condition is valid for all $i$, we have $\varphi_{ij}=\varphi_{j}$, $\psi _{ij} = \psi _i $, and $B_{ij}=B$. By symmetry, $f_{ij} (A _i , A _j )= f _{ji}(A_j,A_i)$, which means $\varphi _i = \psi _i $, $C _{ij} = C _{ji}$, and $B$ is symmetric. We thus have
\[
\frac{\partial  f_{jj}}{\partial A_j }=2\frac{\partial  f_{ij}}{\partial A_i }=2\varphi _{j} +2B(\_\,,A _j )
\]
and therefore $f_{ii}$ are necessarily of the form $f_{ii}(A_i)=C_{ii}+ 2 \varphi _i (A_i)+B(A_i, A_i )$. This proves the statement.$\qquad\blacksquare$

\subsubsection{The generalized Jacobi group consists of generalized quantomorphisms} Let us now assume that $d\geq 2$. Using Darboux coordinates on $V$, we can write  elements in $L(W,V)$ as $ \mathbf{Z} =( \mathbf{Q} , \mathbf{P} )$, 
where $ \mathbf{Q} , \mathbf{P} $ are $k \times m$ matrices, $2k=n$. We observe that all the results obtained in \S\ref{Sec:ILM} regarding the ILM cocycle extend to the present case. It suffices to note that $ \Omega ^W $ is exact, with potential form $ \alpha ^W \in \Omega ^1(L(W,V), \operatorname{Sym}(W))$ given by
\[
\alpha ^W ( \mathbf{Q} , \mathbf{P} )( \dot{\mathbf{Q}} , \dot{\mathbf{P}} )= \frac{1}{2} \left( \mathbf{Q} ^T \dot{ \mathbf{P} }- \mathbf{P} ^T \dot{ \mathbf{Q} }\right) ^{sym}.
\]
Therefore, one can define the ILM cocycle exactly as before, and consider the central extension $\operatorname{Diff}\left( L(W,V),\Omega^W \right)\times _B \operatorname{Sym}(W)$ with cocycle
\[
B(\eta_1,\eta_2)=\int_0^{\eta_2(0)}\left(\eta_1^*\alpha^W -\alpha^W \right).
\]
This central extension is isomorphic to the group $ \operatorname{Aut}\left( L(W,V)\times \operatorname{Sym}(W), \mathbf{d} \alpha ^W- \mathbf{d} S \right) $ of automorphisms of the bundle $L(W,V)\times \operatorname{Sym}(W) \rightarrow L(W,V)$ that preserve the $\operatorname{Sym}(W)$-valued $1$-form $ \alpha ^W- \mathbf{d} S $. This group plays the role of the quantomorphism group when $d \geq 2$. At the infinitesimal level, we have the Lie algebra isomorphisms
\[
\mathfrak{X}\left( L(W,V),\Omega^W \right) \times_C\operatorname{Sym}(W)\rightarrow\mathfrak{aut}\left( L(W,V)\times\operatorname{Sym}(W),\alpha^W -\mathbf{d}e \right) \rightarrow \mathcal{F}_{adm}(L(W,V), \operatorname{Sym}(W)).
\]
This shows, by Lemma \ref{F_adm}, that the group $\operatorname{Diff}\left( L(W,V),\Omega^W \right)\times _B \operatorname{Sym}(W)$ is finite dimensional. Note that the Lie bracket on the space of admissible functions is given by $\{\!\{f,g\}\!\}$, see \eqref{Lie_bracket_d}.

Of course, we still have the same group inclusions as in \S\ref{remarkable_subgroup}, for example $\operatorname{Sym}(W)$, $\operatorname{Sp}(V,\Omega)$, $\operatorname{H}(V,\Omega;W)$, and $ \operatorname{Jac}(V, \Omega ;W)$ are included in the central extended group as, respectively,
\[
A\mapsto (id,A),\quad g\mapsto (g,0),\quad (\mathbf{V},S)\mapsto (\tau_\mathbf{V},S),\quad (g,\mathbf{V},A)\mapsto (\tau_\mathbf{V}\circ g,A),
\]
where $\tau_\mathbf{V} (\mathbf{Z} ):=\mathbf{Z} +\mathbf{V} $ is the translation. However, since the generalized Jacobi group and the ILM central extended group have the same dimension, we deduce that they are equal, being connected. Taking the tangent map to this Lie group isomorphism and composing with the Lie algebra isomorphism $ \Gamma $ defined in Lemma \ref{lemma:LAisomorphism_d}, we get the explicit identification
\[
\operatorname{Sym}(V\oplus W)\simeq
\left(\operatorname{Sym}(V)\,\circledS\,L(V,W^*)\right)\times
\operatorname{Sym}(W)\to\mathcal{F}_{adm}(L(W,V),\operatorname{Sym}(W)),
\]
\[
(S,A,B)\mapsto \left(
\frac{1}{2}\,\mathbf{Z}^TS\mathbf{Z}+(A\mathbf{Z})^{sym}+B\right)=f( \mathbf{Z} ),
\]
between the Bloch-Iserles variables and the space of admissible functions. 

\section{Conclusions and open questions}

This paper showed that the Bloch-Iserles system on the space of symmetric matrices is a geodesic flow on the Jacobi group. This result arose essentially from the observation that the BI system possesses a Vlasov moment formulation, which in turn allows to identify momentum map solutions arising from the Klimontovich map.  The Lie-Poisson structure of the general BI system was studied in detail and a special type of momentum map was used to transform BI on the Jacobi group $\operatorname{Sp}(V,\Omega) \,\circledS\, \operatorname{H}(V,\Omega;W)$ into a system on the direct product group $\operatorname{Sp}(V,\Omega) \times H(V,\Omega,W)$. 

Moreover, this paper identified for the first time Vlasov-type systems with coadjoint orbits of the ILM group. Then, the Vlasov formulation of BI dynamics was shown to arise from a deep geometric construction, in which the Jacobi group underlying BI dynamics emerges as a subgroup of the quantomorphism (ILM) group. When trying to extend this result to the generalized Jacobi group, we found that the latter produces a natural extension of the quantomorphism group, so that the symplectic form may take values in the space of (invertible) symmetric matrices. This result was deeply inspired by the recent progress in \cite{Vi2010}. 

Many open questions emerge in the present work. Probably, the most intriguing  concerns moments and its underlying Lie group. In \cite{ScWe1994}, the authors mention (without giving any detail) that this Lie group is given by the jet group associated to quantomorphisms. However, this statement was never considered in later developments and this poses an important question concerning moment dynamics, which would then be explained as coadjoint dynamics on a group of jets. The same questions emerge for kinetic moments in \S\ref{Sec:IntMomClo}. In this case, the Lie algebra properties associated to kinetic moments (dual to differential operators) \cite{GiHoTr2008,HoTr2009} are known to possess no underlying Lie group. Still, given the identification of kinetic moments with fiberwise polynomials, it is natural to wonder whether kinetic moments may allow for a geometric description in terms of a groupoid involving jets. However, this remains an outstanding open question that we plan to pursue in future work.

\appendix

\section{Appendix}

\subsection{Lie-Poisson and Euler-Poincar\'e equations}\label{LP+EP}
Recall that the $\pm$ Lie-Poisson bracket associated to a Lie
algebra $\mathfrak{g}$ is
\[
\{f,g\}_{\pm}(\mu)=\pm\left\langle\mu,\left[\frac{\delta f}{\delta
\mu},\frac{\delta f}{\delta \mu}\right]\right\rangle,\qquad
\mu\in\mathfrak{g}^*,\quad f,g\in\mathcal{F}(\mathfrak{g}^*),
\]
where $\langle\,,\rangle$ denotes the pairing between the Lie
algebra and its dual. Given a Hamiltonian $h$ on $\mathfrak{g}^*$,
the associated equations of motion are thus given by
\begin{equation}\label{LP}
\dot\mu=\mp\operatorname{ad}^*_{\frac{\delta h}{\delta\mu}}\mu.
\end{equation}
These equations are obtained by reduction of the canonical Hamilton's
equations associated to a $G$-invariant Hamiltonian on the phase
space $T^*G$. In the formulas above, the upper sign corresponds to
right invariance and the other to left invariance. When the
Hamiltonian $h(\mu)$ is regular (this is not the case for the Vlasov
equation), the Legendre transform $\xi=\delta h/\delta\mu$ takes to
the Lagrangian side, thereby yielding the \textit{Euler-Poincar\'e
equations}
\begin{equation}\label{EP}
\frac{d}{dt}\frac{\delta\ell}{\delta\xi}=\mp\operatorname{ad}^*_{\xi}\frac{\delta\ell}{\delta\xi}
\end{equation}
relative to a Lagrangian $\ell:\mathfrak{g}\rightarrow\mathbb{R}$.
These equations are also obtained by reduction of Hamilton's
variational principle, associated to a $G$-invariant Lagrangian on
$TG$. An Euler-Poincar\'e approach to the Vlasov system has been
proposed in \cite{CeHoHoMa98}. Throughout this paper we shall
consider both the Lie-Poisson and Euler-Poincar\'e settings whenever
the Legendre transform can be carried out.

\subsection{Notation for symplectic vector spaces}\label{notation}

 The symplectic form
$\Omega:V\times V\to\Bbb{R}$ generates the usual isomorphism
\[
\Omega^\flat:V\rightarrow
V^*\,,\qquad\langle\Omega^\flat(u),v\rangle:=\Omega(u,v)\,,
\]
whose inverse $\Omega^\sharp:=(\Omega^\flat)^{-1}$ is evidently
given by
\[
\Omega^\sharp:V^*\rightarrow
V\,,\qquad\Omega(\Omega^\sharp(\alpha),v)=\langle\alpha,v\rangle\,.
\]
In addition, the symplectic form generates the Poisson tensor
\[
\Pi:V^*\times V^*\rightarrow\mathbb{R}\,,\qquad
\Pi(\alpha,\beta):=\Omega(\Omega^\sharp(\alpha),\Omega^\sharp(\beta))=\langle\alpha,\Omega^\sharp(\beta)\rangle\,,
\]
which yields its associated Poisson bracket
\[
\{f,g\}(v)=\Pi(\mathbf{d}f(v),\mathbf{d}g(v)).
\]
Upon defining
\[
\Pi^\sharp:V^*\rightarrow V\,,\qquad
\Pi(\alpha,\beta):=\langle\alpha,\Pi^\sharp(\beta)\rangle\,,
\]
we obtain the relation
\[
\Pi^\sharp=\Omega^\sharp\,.
\]
The corresponding matrix relations among the symplectic form and its
 Poisson tensor may be easily found by fixing a basis
on $V$, so that $V^*$ is endowed with the corresponding dual basis.
We have
\[
[\Pi]=[\Pi^\sharp]=[\Omega^\sharp]=[\Omega^\flat]^{-1}=-[\Omega]^{-1}\,,
\]
where the square bracket notation indicates the associated matrix.
Choosing a basis on $V$, we can write $V=\mathbb{R}^{2k}$ and
\[
[\Omega]=\mathbb{J},\quad [\Pi]=\bar{\mathbb{N}}
\]
with $\bar{\mathbb{N}}=-\mathbb{J}^{-1}$.

\subsection{Semidirect-product reduction by stages}\label{RedByStages} \normalfont
Consider a semidirect product group $G\,\circledS\,H$ with group multiplication
\[
(g,h)(g',h')=(gg',h(g\cdot h')),
\]
where $g\cdot h$ denote the action of $G$ on $H$ by group homomorphisms. This paper considers the case $G=\operatorname{Sp}(\mathbb{R}^{2k},-\bar{\mathbb{N}}^{-1})$ and $H=\operatorname{H}(\mathbb{R}^{2k},-\bar{\mathbb{N}}^{-1};\mathbb{R}^d)$. The (\textit{left}) Lie-Poisson structure on $(\mathfrak{g}\,\circledS\,\mathfrak{h})^*$ is obtained by reduction of the canonical cotangent bundle $T^*(G\,\circledS\,H)$ by the whole group $G\,\circledS\,H$. This is the usual Lie-Poisson reduction. Note that this approach does not take into account of the particular semidirect product structure of the group.

However, one can also carry out the Poisson reduction in \textit{two stages}, by making concrete use of the peculiar form of the group. First we reduce the cotangent bundle by the normal subgroup $H$ of $G\,\circledS\,H$ and we get the intermediate Poisson manifold
\[
T^*(G\,\circledS\,H)/H=T^*G\times\mathfrak{h}^*
\]
by the quotient map $(g,\dot g,h,\dot h)\mapsto (g,\dot g,h^{-1}\dot h)=(g,\dot g,\eta)$. The Poisson structure on $T^*G\times\mathfrak{h}^*$ is the direct sum of the canonical Poisson structure on $T^*G$ and the Lie-Poisson structure on $\mathfrak{h}^*$. Then we reduce the Poisson manifold $T^*G\times\mathfrak{h}^*$ by $G$ and obtain the dual semidirect product Lie algebra
\[
(T^*G\times\mathfrak{h}^*)/G=(\mathfrak{g}\,\circledS\,\mathfrak{h})^*
\]
from the quotient map $(g,\dot g,\eta)\mapsto (g^{-1}\cdot g,g^{-1}\cdot \eta)$. By the general result of Poisson reduction by stages (\cite{MaMiOrPeRa2007}), the Poisson bracket obtained on $(\mathfrak{g}\,\circledS\,\mathfrak{h})^*$ by the reduction in two steps coincides with the Poisson bracket obtained by reduction in one step
by the whole group $G\,\circledS\,H$, that is, the Lie-Poisson bracket.

\subsection{Untangling Poisson brackets}\label{untangling-momap-appendix}

Let $(P,\{\cdot,\cdot\}_P)$ be an arbitrary Poisson manifold and $G$ a Lie group acting on the \textit{left\/} on $P$ by Poisson diffeomorphisms. Consider the Poisson manifold $T^*G\times P$ endowed with the sum of the canonical Poisson bracket on $T^*G$ and the Poisson structure $\{\cdot,\cdot\}_P$ on $P$. Consider the diagonal action of $G$ on $T^*G\times P$ induced by the cotangent lift of left translation on $G$ and the $G$ action on $P$. As shown in \cite{KrMa1987}, the reduced bracket on $(T^*G\times P)/G\simeq \mathfrak{g}^*\times P$ obtained by Poisson reduction is given by
\begin{equation}\label{genLP-BI}
\{f,g\}(\mu,m)=-\left\langle\mu,\,\left[\frac{\delta f}{\delta
\mu},\frac{\delta g}{\delta
\mu}\right]\right\rangle+ \left\{f, g\right\}_P
+\left\langle\mathbf{d}_{m\,} g,\left(\frac {\delta f}{\delta
\mu}\right)_{\!\!P\!\!}(m)\right\rangle -\left\langle\mathbf{d}_{m\,}
f,\left(\frac {\delta g}{\delta
\mu}\right)_{\!\!P\!\!}(m)\right\rangle,
\end{equation}
where the vector field $\xi_P$ on $P$ denotes the Lie algebra
action of $\xi\in\mathfrak{g}$ on $P$. We will refer to brackets of the form \eqref{genLP-BI} as \textit{generalized Lie-Poisson structures}.

This type of Poisson structure arises in several physical contexts, such as the dynamics of charged fluids, superfluids and liquid crystals \cite{Ho1987,HoKu1982,GBTr2010, GBRa2011}. The remarkable fact about generalized Lie-Poisson structures is the existence of an \emph{untangling map}, that is a Poisson isomorphism which takes the bracket \eqref{genLP-BI} into a direct-product structure which does not explicitly involve the $\mathfrak{g}-$action on $P$. Rather,  this action is implicitly contained in the new dynamical variables. The only hypothesis for the existence of the untangling map is that the $G$-action on $P$ allows for an equivariant momentum map \cite{KrMa1987}. Indeed, we recall the following general result

\begin{proposition}[\cite{KrMa1987}]\label{untangling_prop}
Let $G$ be a Lie group acting on a Poisson manifold $P$ from the left.
Suppose the $G$-action on $P$ admits an equivariant momentum map $\mathbf{J}_P:P\rightarrow\mathfrak{g}^*$
and define the following Poisson structure on $\mathfrak{g}^*\times
P$
\[
\left\{f,g\right\}_{u}= -\left\langle\mu,\left[\frac{\delta
f}{\delta\mu},\frac{\delta
g}{\delta\mu}\right]\right\rangle+\left\{f,g\right\}_P,
\]
Then the map
\begin{equation}\label{untangling_map}
u:\left( \mathfrak{g}^*\times P\,,\left\{\cdot,\cdot\right\}
\right) \,\to \left(\mathfrak{g}^*\times
P\,,\left\{\cdot,\cdot\right\}_u\right),\quad (\mu,m)\,\mapsto\left(\mu+\mathbf{J}_P(m)\,,m\right)
\end{equation} 
is a Poisson diffeomorphism.
\end{proposition}
The name ``untangling map'' is due to \cite{Ho1987}, who
showed how this map applies to electromagnetic charged fluids. In the present context, the special form of the
Bloch-Iserles system allows us to untangle its bracket. Indeed, it suffices to consider the case 
\[
T^*G \times P= T^* \operatorname{Sp}( \mathbb{R}^{2k}, \mathbb{J} ) \times \mathfrak{h}^*(\R^{2k},\mathbb{J};\R^d)\,.
\]
As explained in Appendix \ref{RedByStages} with the help of reduction by stages, the reduced bracket \eqref{genLP-BI} in this case recovers the Lie-Poisson bracket on $\mathfrak{jac}(\R^{2k},-\Bbb{N}^{-1};\R^d)^*$.

\subsection{Alternative form of the Bloch-Iserles equations}\label{UntangledBI}

This Appendix shows the explicit form of the BI equations under the untangling map, that is under the change of variable\
\[
\mathcal{Y}:=\mathcal{X}+\frac12\,\bar{\Bbb{N}}\,\mathcal{K}^T\mathcal{M}^{-1}\mathcal{K}\,.
\]
One starts by writing the Bloch-Iserles Hamiltonian as
\begin{align*}
h(\mathcal{Y,K,M})
=\ &
\frac12\operatorname{Tr}\!\left(
\left(\mathcal{Y}+\frac12\bar{\Bbb{N}}\,\mathcal{K}^T \mathcal{M}^{-1}\mathcal{K}\right)
(-\bar{\Bbb{N}}^2)
\left(\mathcal{Y}+\frac12\bar{\Bbb{N}}\,\mathcal{K}^T \mathcal{M}^{-1}\mathcal{K}\right)^{\!\!T\,}
\right)
\\
&+
\frac{1}{4}\operatorname{Tr}\!\left(\mathcal{K}\,(-\bar{\mathbb{N}}^2)\,\mathcal{K}^T\right)
+
\frac{1}{8}\operatorname{Tr}\!\left(\mathcal{M}^2\right)
\\
=\ &
\frac12\operatorname{Tr}\!\left((\mathcal{Y}\,\bar{\Bbb{N}})^2\right)
-
\frac12\operatorname{Tr}\!\left(\mathcal{Y}\,(-\bar{\Bbb{N}}^2)\,\mathcal{K}^T \mathcal{M}^{-1}\mathcal{K}\,\bar{\Bbb{N}}\right)
-
\frac18\operatorname{Tr}\!\left(\!\Big((-\bar{\Bbb{N}}^2)\,\mathcal{K}^T \mathcal{M}^{-1}\mathcal{K}\Big)^{\!2}\right)
\\
&+
\frac{1}{4}\operatorname{Tr}\!\left(\mathcal{K}\,(-\bar{\mathbb{N}}^2)\,\mathcal{K}^T\right)
+
\frac{1}{8}\operatorname{Tr}\!\left(\mathcal{M}^2\right).
\end{align*}
Upon using the formula
\[
\delta\!\left(\mathcal{K}^T \mathcal{M}^{-1}\mathcal{K}\right)
=
2\left(\mathcal{K}^T \mathcal{M}^{-1}\delta\mathcal{K}\right)^\sym
-
\mathcal{K}^T \mathcal{M}^{-1}\delta\mathcal{M}\mathcal{M}^{-1}\mathcal{K}
\,,
\]
one finds the explicit expression for the derivatives
\begin{align*}
\frac{\delta h}{\delta\mathcal{Y}}
&=
\bar{\Bbb{N}}\mathcal{Y}\bar{\Bbb{N}}
+
\frac12\bar{\Bbb{N}}^2\mathcal{K}^T \mathcal{M}^{-1}\mathcal{K}\bar{\Bbb{N}}
\\
\frac{\delta h}{\delta\mathcal{K}}
&=
\left(\bar{\Bbb{N}}\,\mathcal{Y}\,\bar{\Bbb{N}}^2\right)^\sym
\mathcal{K}^T \mathcal{M}^{-1}
+\frac12\,
\bar{\Bbb{N}}^2
\mathcal{K}^T \mathcal{M}^{-1}\mathcal{K}\,
\bar{\Bbb{N}}^2\mathcal{K}^T \mathcal{M}^{-1}
-
\frac12\,\bar{\Bbb{N}}^2\mathcal{K}^T
\,,
\end{align*}
which, inserted in \eqref{untangled_Hamilton_equations}, produce the untangled Bloch-Iserles system in the form
\begin{equation}
\left\{
\begin{array}{l}
\displaystyle\vspace{0.2cm}\dot{\mathcal{Y}}
=
\left[\mathcal{Y},
\bar{\Bbb{N}}\mathcal{Y}\bar{\Bbb{N}}
\right]
+
\frac12
\left[\mathcal{Y},
\bar{\Bbb{N}}^2\mathcal{K}^T \mathcal{M}^{-1}\mathcal{K}\bar{\Bbb{N}}
\right]\\
\displaystyle\vspace{0.2cm}\dot{\mathcal{K}}=-\mathcal{K}\left(\bar{\Bbb{N}}\mathcal{Y}\bar{\Bbb{N}}^2\right)^\sym\bar{\Bbb{N}}^{-1}-\frac12\,\mathcal{K}\bar{\Bbb{N}}^2\mathcal{K}^T\mathcal{M}^{-1}\mathcal{K}\bar{\Bbb{N}}+\frac12\,\mathcal{M}\mathcal{K}\bar{\Bbb{N}}\\
\dot{\mathcal{M}}=0.
\end{array}
\right.
\end{equation}

\paragraph{Acknowledgments.} We wish to thank Cornelia Vizman for her many valuable comments on the geometry of prequantization Abelian extension. Stimulating discussions with Anthony Bloch, Darryl Holm and Tudor Ratiu are also greatly acknowledged.


\end{document}